\documentclass[a4paper, 11 pt]{article}
\pdfoutput=1
\usepackage{cite}
\usepackage{epsfig,amssymb,bbm,euscript,xspace,xcolor,setspace,amsmath,mathtools,empheq,amsthm,graphicx,mathrsfs,float}
\usepackage[utf8]{inputenc}
\usepackage{hyperref}
\usepackage{bbding}
\usepackage{tikz}
\usepackage{xcolor}
\hypersetup{
    colorlinks=true,
    linkcolor=black,
    citecolor=blue
    } 
\usepackage{tcolorbox}
\usepackage{booktabs}
\usepackage{caption,subcaption} 
\usepackage[belowskip=-20pt,aboveskip=0pt]{caption}
\usepackage[skip=10pt]{caption} 
\usepackage{comment}
\usepackage{wrapfig}
\usepackage[numbers,sort&compress]{natbib}
\usepackage{cite}
\usepackage{makecell}

\begin{document}
\newcommand\blfootnote[1]{%
  \begingroup
  \renewcommand\thefootnote{}\footnote{#1}%
  \addtocounter{footnote}{-1}%
  \endgroup
}

\makeatother

\addtolength{\oddsidemargin}{-1.0cm} 
\addtolength{\textwidth}{2.0cm}
%

\def\g{\gamma}
\def\r{\rho}
\def\s{\sigma}
\def\m{\mu}
\def\n{\nu}
\def\a{\alpha}
\def\e{\epsilon}
\def\k{\kappa}
\def\b{\beta}
\def\d{\delta}
\def\x{\xi}
\def\f{\phi}
\def\p{\pi}
\def\t{\theta}
\def\D{\Delta}
\def\G{\Gamma}
\def\l{\lambda}
\def\pd{\partial}
\def\tq{\tilde{q}}
\def\tf{\tilde{f}}
\def\ta{\tau}
\def\bz{\bar{z}}
\def\mb{\mathcal{B}}
\def\fin{f_{\infty}}
\def\mo{{\mathcal{O}}}
\def\mw{{\mathcal{W}}}
\def\ml{{\mathcal{L}}}
\def\G{\Gamma}
\def\tn{\tilde{n}}
\def\mc{{\mathcal{C}}}
\def\md{{\mathcal{D}}}
\def\bh{\bar{h}}
\def\w{\omega}
\def\ib{{\mathbb{I}}}
\def\tell{\tilde{\ell}}
\def\tr{{\rm tr~}}
\def\spart{S_\ell}
\def\tpart{T_\ell}
\def\be{\begin{equation}}
\def\ee{\end{equation}}
\def\bea{\begin{eqnarray}}
\def\eea{\end{eqnarray}}
\newtheorem*{definition}{Definition}
\newtheorem*{theorem}{Theorem}
\newtheorem*{proposition}{Proposition}
\thispagestyle{empty}
\setcounter{tocdepth}{2}
\baselineskip 10pt
\vspace*{1.0ex}
\newcommand{\pr}[1]{\textcolor{orange}{#1}}

\noindent
\hfill {\small\text{MPP-2026-59}}
\vspace*{2 cm}
\begin{center}
{\Large \bfseries Lecture Notes on Positivity Properties of Scattering Amplitudes}
\end{center}
\rule{\linewidth}{1pt}


\vspace*{2.0ex}

\baselineskip=18pt

{\large \rm  \hspace*{-20 pt} Prashanth Raman$^{a}$} \\
{\small \noindent{\it ${}^{a}$Max-Planck-Institut f\"{u}r Physik,
Boltzmannstr.8,
85748 Garching, Germany.}}\\
\noindent E-mail: \url{praman@mpp.mpg.de}

\centerline{\bf Abstract} \bigskip
 We review completely monotone (CM) and Stieltjes functions, which are classes of functions that obey an infinite hierarchy of positivity constraints. While these are classical mathematical concepts, such properties have recently been shown to arise in many fundamental building blocks and observables of quantum field theory (QFT), such as scalar Feynman integrals in the Euclidean region and Coulomb branch amplitudes in $\mathcal{N}=4$ SYM. 
 
 After reviewing their mathematical structure, we discuss the various physical and geometric origins of these properties—ranging from unitarity and analyticity in scattering amplitudes to the structure of parametric representations in Feynman integrals. We then review several applications, including constraints on the the analytic S-matrix, implications for numerical bootstrap approaches, and connections to positive
geometries, where we present evidence for a close relation between these functions and geometric volume interpretations. These lecture notes are an extended version of lectures give at the {\it Positive Geometry in Scattering Amplitudes and Cosmological Correlators} workshop held at the International Centre for Theoretical Sciences, Bengaluru, in February 2025.

\newpage
\tableofcontents
\onehalfspacing
\newpage

\section{Introduction}

Positivity and convexity are fundamental organizing principles in quantum field theory (QFT), often arising as consequences of basic physical requirements such as unitarity, causality, and analyticity. Unitarity ensures probabilistic consistency, while analyticity encodes causality and locality, leading to powerful constraints on physical observables. These ideas underlie a wide range of developments, including dispersion relations, bounds on effective field theories, and the conformal bootstrap \cite{Arkani-Hamed:2020blm,Bellazzini:2020cot,Simmons-Duffin:2016gjk}.

A modern geometric perspective on these structures has emerged from the study of scattering amplitudes \cite{Arkani-Hamed:2013jha, Arkani-Hamed:2017tmz}. In the framework of positive geometry, amplitudes (or their integrands) are identified with canonical differential forms associated with geometric objects \cite{Arkani-Hamed:2017tmz}. In this approach, positivity is built into the definition, and physical properties such as locality and unitarity arise as consequences of geometry \cite{Herrmann:2022nkh}. In simple cases, amplitudes admit an interpretation as volume-like quantities of dual geometric objects, suggesting a deep connection between geometry and the analytic properties of QFT observables \cite{Hodges:2009hk}.

In perturbative QFT, observables are typically expressed as integrals where the integrand $I$ is often a relatively simple rational function, while the integrated result $A$ is a complicated transcendental function. If integrands are volumes then they are guaranteed to be positive in certain regions. This naturally raises the question of whether positivity properties of the integrand survive integration, and if so, in what form \cite{Arkani-Hamed:2014dca}. Understanding this is important both conceptually and practically, as such properties can provide powerful constraints for analytic and numerical methods.

Recent work has shown that a remarkably strong form of positivity appears in a wide class of fundamental building blocks and observables: complete monotonicity \cite{henn2025positivitypropertiesscatteringamplitudes}. This property imposes an infinite hierarchy of sign constraints on all derivatives of a function and is known to be equivalent to the existence of a representation as a Laplace transform of a positive measure \cite{Choquet}. It therefore provides a bridge between positivity, convexity, and integral representations. A particularly important subclass is given by Stieltjes functions, which satisfy even stronger analytic constraints. They admit a spectral-type representation with a positive measure and possess highly constrained analytic behavior in the complex plane.

Strikingly, both completely monotone and Stieltjes structures arise naturally in QFT \cite{henn2025positivitypropertiesscatteringamplitudes,ditsch2026approximatingfeynmanintegralsusing}. For example, scalar Feynman integrals in the Euclidean region often fall into this class, and similar properties have been observed for Coulomb branch amplitudes in $\mathcal{N}=4$ SYM in suitable kinematic regimes \cite{henn2025positivitypropertiesscatteringamplitudes}. These structures impose non-trivial constraints on the analytic behavior of physical quantities and can be used to extract global information from limited perturbative data. 

From the perspective of positive geometry, the appearance of these properties is natural due to the connection to dual volumes. However, the underlying reasons for their emergence can differ: while in scattering amplitudes these properties are often consequences of unitarity and analyticity, for Feynman integrals they emerge from the structural properties of parametric representations and Symanzik polynomials \cite{henn2025positivitypropertiesscatteringamplitudes,ditsch2026approximatingfeynmanintegralsusing}. Whenever an observable admits a dual volume or Laplace-type representation over a positive domain, complete monotonicity follows immediately.

The aim of these lecture notes is to provide a pedagogical review of these ideas. We will study how general physical principles lead to positivity constraints, how these can be strengthened to complete monotonicity and Stieltjes properties, and how they are used in practice to derive rigorous bounds and to bootstrap non-trivial observables \cite{ditsch2026approximatingfeynmanintegralsusing}. We illustrate these concepts through a range of examples, highlighting both their physical origin and their mathematical structure.

These lecture notes are based on lectures given at the \emph{Positive Geometry in Scattering Amplitudes and Cosmological Correlators} workshop, held at the International Centre for Theoretical Sciences (ICTS), Bengaluru, in February 2025. A recording of the lectures is available at \url{https://www.youtube.com/live/bXsNiIOSZRs?si=1Z2GmDTFIoOU_adk}. 

The notes are organized as follows. In Section~2, we review the mathematical framework of completely monotone and Stieltjes functions. Section~3 surveys examples from quantum field theory where these positivity structures arise. Section~4 discusses applications to the analytic S-matrix, positive geometry, and numerical bootstrap methods. The appendices provide additional background on related positivity classes, the moment problem, and Abelian--Tauberian theorems.
\section{Completely monotone and Stieltjes functions}
\subsection{Completely monotone functions in one variable}

To motivate the discussion that follows, consider the problem of reconstructing a function from its values at positive integers:

Suppose we are given a function $f: \mathbb{R} \rightarrow \mathbb{R}$ specified at all positive integers,
\begin{align}\label{eq:firstprob}
f(n) = a_n, \quad n=1,2,3,\dots
\end{align}
How constraining is this information? In particular:

\begin{itemize}
\item Can we find an interpolating function defined at all real values $x>0$?  
\item Under what conditions is such an interpolating function unique?  
\item Can we construct an analytic continuation of $f$ to complex values of $x$?
\end{itemize}

Clearly, the function is highly undetermined: for example, both $f(x)$ and $f(x)+ \sin(\pi x)$ satisfy the above condition.  

A classic illuminating example is when $a_n = n!$, in which case $f(x) = \Gamma(x+1)$ provides a solution. The celebrated \emph{Bohr--Mollerup theorem} states that the Gamma function is the unique solution among positive functions satisfying
\begin{enumerate}
\item $\log f(x+1) - \log f(x) = \log (1+x)$, \quad $\forall x>0$,
\item Logarithmic convexity: $\log f(x)$ is convex i.e., $(\log f(x))''\,\ge 0$.
\end{enumerate}

Let us examine these conditions more closely. The first condition can be expressed using the forward difference operator $\Delta$,
\begin{align}
\Delta f(x) &= f(x+1) - f(x), \\
\Delta^k f(x) &= \Delta^{k-1} f(x+1) - \Delta^{k-1} f(x), \quad k\ge 2,
\end{align}
so that
\begin{align}
(-1)^{k+1}~\Delta^k \log f(n) \ge 0, \quad \forall k \ge 1, \ n\ge 1.
\end{align}
This is like a discrete derivative, and all such differences have fixed signs.  

The second condition ensures smoothness and imposes a constraint on the derivatives of $f(x)$:
\begin{align}\label{eqref:logconvexity}
f(x)\, f''(x) - \big(f'(x)\big)^2 \ge 0,
\end{align}
which is the \emph{log-convexity} condition.

While the Bohr--Mollerup theorem guarantees uniqueness among positive functions on $(0,\infty)$, if we additionally want to extend the function to complex values, further conditions are necessary. In particular, if $f$ is assumed to be analytic in the right half-plane $\text{Re}(x) > 0$ (as is natural for the Gamma function, which has poles on the negative real axis), then \emph{Carlson's theorem} provides a uniqueness result: a function that is analytic in the right half-plane, of exponential type, and bounded along the imaginary axis is uniquely determined by its values at the positive integers. These growth and analyticity constraints ensure that no “oscillatory” solutions (like $f(x)+ \sin(\pi x)$) can appear in the complex extension.

This example illustrates that imposing natural constraints—such as positivity, convexity, sign constraints on the derivatives and controlled growth—can guarantee the uniqueness of an interpolating function. It also motivates the study of \emph{completely monotone functions}, which generalize these ideas to sequences and functions with higher-order sign constraints.
\subsubsection{Definition and basic properties}
We denote by $f^{(n)}(x) = \frac{d^n}{dx^n} f(x)$ the $n$-th derivative of a function $f$.
\begin{definition}
A function $f\in C^\infty(R)$, $R\subset \mathbb{R}$, is called 
\emph{completely monotone} (CM) if
\begin{align}\label{eq:def:monotonex}
(-1)^n f^{(n)}(x) \ge 0 \,,\quad n\ge 0\,,\ \forall x\in R\,,
\end{align}
and \emph{absolutely monotone} (AM) if $f^{(n)}(x)\ge 0$ for all $n\ge 0$ and all $x\in R$.
\end{definition}
\begin{figure}[H]
    \centering
 \includegraphics[width=0.65\linewidth]{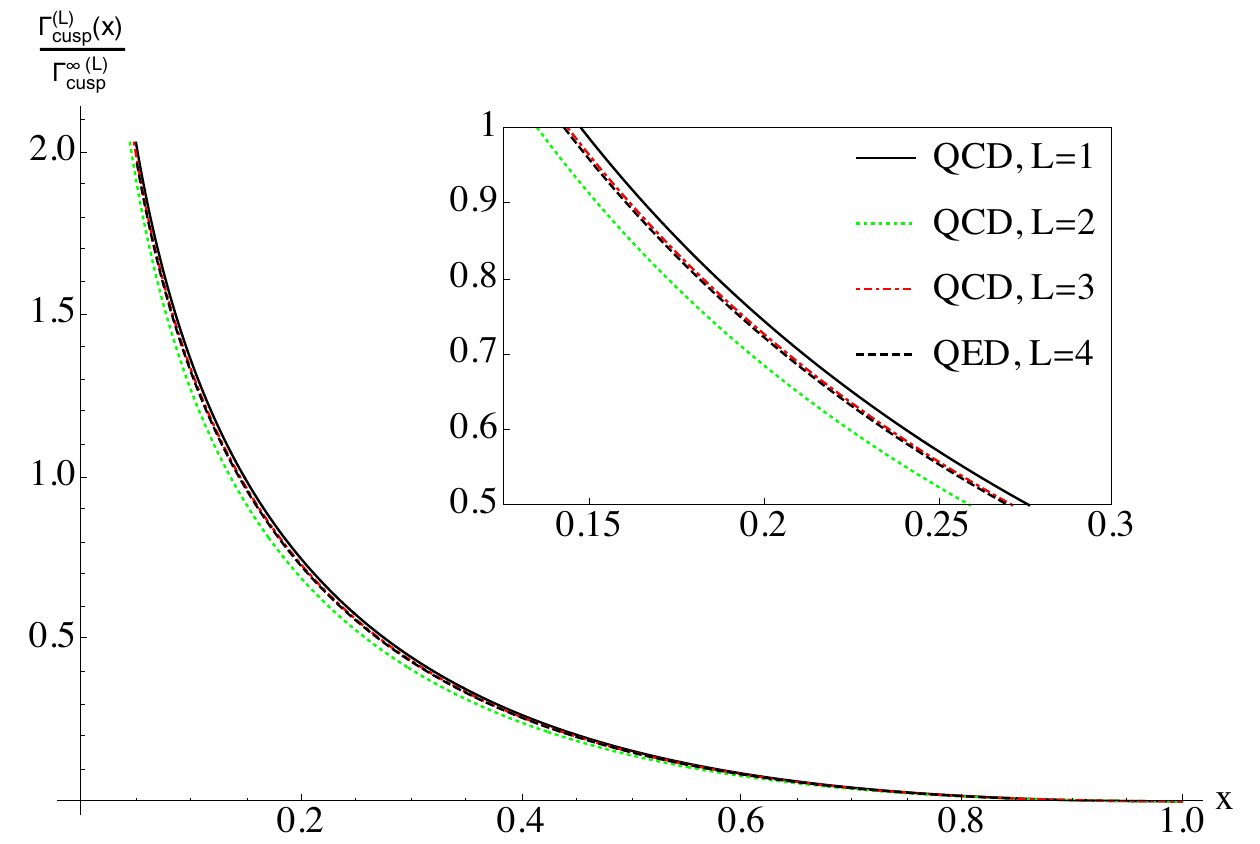}
    \caption{Plot of the angle-dependent cusp anomalous dimension for QCD and QED based on available perturbative data from ref~\cite{henn2025positivitypropertiesscatteringamplitudes}. As we will see in Section~\ref{CAD} that this quantity is expected to be  CM on $(0,1)$.}
    \label{fig:CM_Plot}
\end{figure}
\vspace*{10 pt}
It follows immediately from the definition that if $f$ is CM/AM on $R=(a,b)$, then $g(x)=f(-x)$ is respectively AM/CM on $-R=(-b,-a)$. Furthermore, if $a+b<\infty$ and $h(x)=f(a+b-x)$, then
\begin{equation}\label{eq:shiftabx}
h^{(n)}(x)=(-1)^n f^{(n)}(a+b-x)\,,    
\end{equation}
so complete and absolute monotonicity are closely related. In the following, we focus exclusively on completely monotone functions.

The condition in eq.~\eqref{eq:def:monotonex} imposes infinitely many constraints. In particular, the cases $n=0,1,2$ imply that CM functions are non-negative, monotonically decreasing, and convex on $R$. Consequently, CM functions and their derivatives exhibit smooth, featureless decay, as illustrated in Fig.~\ref{fig:CM_Plot}.

\paragraph{Examples.}
\begin{enumerate}
\item $\frac{1}{x+\alpha}$ with $\alpha>0$, is CM on $(0,\infty)$.
\item $\beta^x$ with $0<\beta<1$, is CM on $(0,\infty)$.
\item $-\log x$ is CM on $(0,1)$.
\end{enumerate}

\paragraph{Basic properties.}
Using the product and chain rule for derivatives, one can establish several structural properties of CM functions. We list them here and refer to 
\cite{WidderWidder+2015,merkle2012completelymonotonefunctions} for proofs. 
Let $f,g$ be CM functions on $(a,b)$.
\begin{enumerate}
    \item \textbf{Convex cone:} For $\lambda,\mu\ge 0$, the combination
    \begin{equation}
       \lambda \,f(x)+\mu\, g(x)   
    \end{equation}
    is CM. Thus, CM functions form a convex cone.

    \item \textbf{Closure under products:} The product $f(x)\,g(x)$ is CM.

    \item \textbf{Closure under derivatives:} For every $n\in\mathbb{N}$, the function
    \begin{equation}
    (-1)^n \,f^{(n)}(x)
    \end{equation}
    is CM.

    \item \textbf{Closure under limits:} If $\{f_n\}$ is a sequence of CM functions converging pointwise to $f$, then $f$ is CM.

    For example, $f_n(x)=(1-\tfrac{x}{n})^n$ is CM on $(0,n)$ and converges pointwise to $e^{-x}$ as $n\to\infty$, which is therefore CM on $(0,\infty)$.

    \item \textbf{Composition (I):} Let $h:(a,b)\to(a,b)$ be a \emph{Bernstein function}, i.e.\ $h\ge 0$ and $h'$ is CM. Then $f\circ h$ is CM on $(a,b)$.

    In particular, if $f,g$ are CM, then
    \begin{equation}
    f\!\left(\int_0^x g(t)\,dt\right)
   \end{equation}
    is CM (on any interval where the argument lies in $(a,b)$).

    \item \textbf{Composition (II):} Let $h:(a,b)\to\mathbb{R}$ be absolutely monotone. If $f:(a,b)\to(a,b)$ is CM, then $h\circ f$ is CM on $(a,b)$.

    For example, if $\log f(x)$ is CM on $(0,\infty)$, then $f(x)=e^{\log f(x)}$ is also CM.
 \item \textbf {Truncation of the Taylor series:} Using the Taylor's reminder theorem  
 \begin{align}
  f(x)&= \sum\limits_{k=0}^{m-1} \frac{(x-x_0)^k}{k!} f^{(k)}(x_0) + \int_{x_0}^x du \frac{(x-u)^{m-1}}{(m-1)!} f^{(m)}(u).
 \end{align}
 and noting that the CM property implies that the reminder term has fixed sign. It follows that the truncation of the Taylor expansion of any CM function $f(x)$, around any point $x=x_0 \in R$ to even and odd order, provides a lower and upper bound for the function respectively i.e.,
\begin{equation}
\sum\limits_{k=0}^{2m} \frac{(x-x_0)^k}{k!} f^{(k)}(x_0) \le f(x) \le \sum\limits_{k=0}^{2m-1} \frac{(x-x_0)^k}{k!} f^{(k)}(x_0) ~{\rm~for}~ x>x_0.
  \end{equation}
  
\end{enumerate}
An important feature of completely monotone functions is their strong regularity: they are real-analytic on $(a,b)$ and, in fact, admit analytic continuations beyond their domain \cite{Bernstein:1926:Lecons,boas1}.
\begin{theorem}
\textbf{Bernstein's little theorem.} \label{theorem:BLT}
If $f$ is completely monotone on $(a,b)$, then it admits an analytic continuation to a disk of radius $(b-a)$ centered at $x=b$ (after suitable translation of the interval).
\end{theorem}
If $(a,\, b)=(0,\,\infty)$ then by the above result CM functions $f(z)$ should admit an analytic extension to the entire right half plane i.e., $\Re(z)\, >0$. This is indeed true and  the Bernstein-Hausdorff-Widder theorem \cite{widder1941laplace} states that $f$ is CM on $(0,\infty)$ iff it is the Laplace transform of a positive function \footnote{The theorem is usually stated with a positive non-decreasing Borel measure $d\nu$ as measures are in general allowed to be distributional. This is equivalent eq.~\eqref{eq:BHW} if the measure admits a well defined density $d\nu(t) = \mu(t)\,dt $ with $\mu(t) \ge 0$ which we have assumed here.}. The Laplace transform in eq.~\eqref{eq:BHW} provides the analytic continuation  as it converges for any $x=x_1+ i\, x_2$ with $x_2 \in \mathbb{R}$ whenever $x_1>0$ and therefore converges in the right half plane.
\noindent 
\begin{tcolorbox}
\begin{theorem}\textnormal{\textbf{Bernstein-Hausdorff-Widder~(BHW)}}\\
A function $f(x)$ is completely monotonic on $(0,\infty)$ if and only if it admits an integral representation 
\begin{equation} \label{eq:BHW}
    f(x) = \int_0^\infty dt \; e^{-xt} \mu(t), \text{     with  }\mu\geq 0.
\end{equation}
\end{theorem}
\end{tcolorbox}\noindent
The Bernstein--Hausdorff--Widder (BHW) theorem provides a characterization of CM functions on $(0,\infty)$ and can be used to construct non-trivial examples. For instance,
\begin{align}
f(x)=\frac{\log x}{x-1}, 
\quad \text{with} \quad 
\mu(t) = \int_0^{\infty} \frac{e^{-t y}}{y+1}\,dy,
\end{align}
which is manifestly positive.

However, this characterization is not well-suited for testing whether a given function is CM, since it requires computing the inverse Laplace transform,
\begin{align}
\mu(t) = \mathcal{L}^{-1}(f)(t) = \frac{1}{2\pi i}\int_{\gamma-i \infty}^{\gamma+i \infty} dx \, e^{tx} f(x),
\end{align}
where $\gamma \in \mathbb{R}$ is chosen so that the integral converges. This is typically intractable analytically for generic functions, and while numerical inversion is possible, certifying complete monotonicity requires verifying $\mu(t)\ge 0$ for all $t>0$.

Nevertheless, partial information about $\mu(t)$ can be extracted from the asymptotic behavior of $f(x)$: the behavior near $x\to\infty$ and $x\to 0$ constrains $\mu(t)$ near $t\to 0$ and $t\to\infty$, respectively, via Tauberian theorems (see appendix~\ref{app:tauberian}).
\subsubsection{Convexity properties}
Another important feature of completely monotone functions is that they satisfy several remarkable convexity properties. These lead to non-trivial constraints on the function and its derivatives, and are useful in numerical bootstrap approaches.

These properties can be viewed as successive strengthenings of ordinary convexity: from additive inequalities, to quadratic constraints on derivatives, to multiplicative relations. We summarize some of these below from references~\cite{Fink1982KolmogorovLandau, widder1941laplace,Kimberling1974Probabilistic,NiculescuSra2023HornichHlawka}.

A function $f(x)$ is convex in a region $R$ if for any $x,y \in R$ and $0<t<1$,
\begin{align}
f(t\,x+(1-t)\,y) \le t\, f(x) + (1-t)\,f(y).
\end{align}
CM functions obey a stronger generalization known as the Hornich--Hlawka property.

\begin{enumerate}

\item \textbf{Hornich--Hlawka property:} If $f(x)$ is CM on $(0,\infty)$, then for any $x_1,\dots,x_n \ge 0$,
\begin{align}
\sum_{i=1}^n f(x_i) - \sum_{i<j} f(x_i+x_j) + \cdots + (-1)^{n-1} f\!\left(\sum_{i=1}^n x_i\right) \ge 0.
\end{align}
This can be viewed as a multi-point additive generalization of convexity. It follows from
\begin{align}
1 - \prod_{i=1}^n (1 - e^{-y\, x_i}) \ge 0,
\end{align}
by integrating against a positive measure $\mu(y)$.

\item \textbf{Hankel positivity:} A function $f(x)$ is CM on $(0,\, \infty)$, if and only if the Hankel matrices
\begin{align}
(H_0)_{i,j} = (-1)^{i+j} f^{(i+j)}(x), 
\quad
(H_1)_{i,j} = (-1)^{i+j+1} f^{(i+j+1)}(x), \quad i,j=0,\dots,n
\end{align}
are positive semidefinite for all $n \ge 0$ and $x>0$.

For $\xi \in \mathbb{R}^{n+1}$ and $l \ge 0$, consider
\begin{align}
I = \sum_{i,j=0}^n (-1)^{i+j+l} f^{(i+j+l)}(x)\,\xi_i \xi_j.
\end{align}
Using $f(x)=\int_0^\infty e^{-xt}\mu(t)\,dt$, we obtain
\begin{align}
I = \int_0^\infty dt\, \mu(t)\, e^{-xt}\, t^l \left(\sum_{i=0}^n t^i \xi_i \right)^2 \ge 0.
\end{align}
Thus, the quadratic form is non-negative, and hence the associated matrix is positive semidefinite. Choosing $l=0,1$ yields the result.

This provides a refinement of convexity at the level of derivatives: instead of constraining $f$ itself, we demand positivity of quadratic forms built from its higher derivatives. 

Conversely, positivity of these Hankel matrices implies the non-negativity of all signed derivatives $(-1)^n f^{(n)}(x)$, since their diagonal entries are non-negative. Together with regularity conditions, this provides a characterization of complete monotonicity \cite{widder1941laplace}.

This yields an infinite set of quadratic constraints on derivatives. For instance,
\begin{align}
\begin{vmatrix}
f(x) & -f'(x) \\ 
-f'(x) & f''(x)
\end{vmatrix} \ge 0
\end{align}
implies the \emph{log-convexity} condition we saw in eq.~\eqref{eqref:logconvexity} with higher-order constraints arising from larger determinants.

\item \textbf{Schur convexity:}  Let $m_1 \ge \dots \ge m_n$ be non-negative integers. Define
\begin{align}
u_x(m_1,\dots,m_n) = (-1)^{m_1} f^{(m_1)}(x) \cdots (-1)^{m_n} f^{(m_n)}(x).
\end{align}
Then $u_x(m_1,\dots,m_n)$ is monotone under redistribution of the $m_i$ that preserves their sum and ordering in the sense that
\begin{align}
\sum_{i=1}^k m_i \le \sum_{i=1}^k m_i' \quad \text{for all } k=1,\dots,n-1,
\end{align}
and
\begin{align}
\sum_{i=1}^n m_i = \sum_{i=1}^n m_i' \quad \implies \quad u_x(m_1,\dots,m_n) \le u_x(m_1',\dots,m_n').
\end{align}

This can be viewed as a multiplicative strengthening of convexity: instead of linear or quadratic combinations, one considers ordered redistributions of derivative indices, and demands monotonicity under majorization.
Some examples (see \cite{merkle2012completelymonotonefunctions}):

\noindent (a) For $m=(1,1,1)$ and $m'=(2,1,0)$, Schur convexity implies
\begin{align}
\big(f'(x)\big)^3 \ge f''(x)\, f(x).
\end{align}
This inequality does not follow from Hankel matrix positivity, and illustrates that Schur convexity imposes constraints beyond those captured by Hankel determinants.

\noindent (b) For any integers $k,j \ge 0$, the function
\begin{align}
x \;\mapsto\; \bigg|\frac{f^{(k+j)}(x)}{f^{(k)} (x)} \bigg|
\end{align}
is decreasing.

\end{enumerate}

\medskip

\noindent
In summary, these properties form a hierarchy of increasingly strong constraints:
Hornich--Hlawka encodes global additive inequalities generalizing convexity; Hankel matrix positivity refines this to quadratic constraints on derivatives (including log-convexity); and Schur convexity further imposes higher-order multiplicative inequalities that are not captured by finite Hankel conditions.

\subsubsection{Interpolation of sequences via completely monotone functions}

Let us now return to the question posed at the beginning of this section and see how CM functions can be used to solve the interpolation problem in eq.~\eqref{eq:firstprob}. A useful result in this context is the following:

\begin{theorem}[Müntz theorem for CM functions, \cite{5b5486ec-e7c3-38ae-810e-8390daa67e61}]
Let $(a_n)_{n\in \mathbb{N}}$ be a strictly increasing sequence of positive numbers, $0<a_1<a_2<\dots$, with $ \lim\limits_{{k \rightarrow \infty}}a_k \rightarrow \infty$  such that
\begin{align}
\sum_{n=1}^\infty \frac{1}{a_n} = \infty.
\end{align}
Suppose $f$ and $g$ are completely monotone functions on $(0,\infty)$ satisfying
\begin{align}
f(a_n) = g(a_n), \quad \forall n \in \mathbb{N}.
\end{align}
Then
\begin{align}
f(x) = g(x), \quad \forall x \in (0,\infty).
\end{align}
\end{theorem}

In other words, if you know a CM function on $(0,\infty)$ on a set of points that don't thin out too quickly, then it is uniquely determined. In particular, if we find a CM function with prescribed values at the positive integers, then it is unique.

\medskip
Rather remarkably, if such a function exists, it can be constructed explicitly using the Newton series:

\begin{theorem}[Theorem 3.1 in \cite{LambyMarichalZenaidi2017}]
Let $I \subset \mathbb{R}$ be a right-unbounded open interval, and let $f : I \to \mathbb{R}$ be infinitely differentiable. Suppose that for some $q \in \mathbb{N}$, the derivative $f^{(q)}(x)$ is completely monotone on $I$. Then, for any $a \in I$, $f$ admits the Newton series expansion
\begin{align}\label{eq:newtonseries}
f(x) = \sum_{k=0}^{\infty} \binom{x-a}{k} \, \Delta^k f(a), \quad x \in I,
\end{align}
and the series converges uniformly on compact subsets of $I$.
\end{theorem}

The above theorem shows that it is sufficient for some derivative of the function to be CM. This explains, for instance, why the Bohr--Mollerup characterization works: even though $f(x) = \log \Gamma(1+x)$ is not CM, one of its derivatives, the trigamma function,
\begin{equation}
f''(z) = \int_0^\infty dt~ \frac{t\, e^{-t}}{1-e^{-t}}~ e^{-t z},
\end{equation}
is CM. Using eq.~\eqref{eq:newtonseries} with $a=1$ and $f(n) = \log (n-1)!$ for $n=1,2,\dots$, we obtain the Newton series
\begin{equation}
f(z) = \sum_{n=1}^{\infty} \binom{z}{n} \sum_{k=1}^n (-1)^{n-k} \binom{n-1}{k-1} \log k, \qquad \Re(z) > 0,
\end{equation}
which is known as the \emph{Stern series} for the log-gamma function.

\medskip
We can repeat this construction for other sequences. For example, consider
\begin{align}
a_n = \frac{1}{x_0+n}, \quad x_0>0,
\end{align}
which is obtained by sampling the function $f(x) = 1/x$ at $x = x_0+n$, $n\in \mathbb{N}_0$. The Newton series then yields
\begin{equation}
\frac{1}{x} = \frac{1}{x_0} \sum_{k=0}^{\infty} (-1)^k \frac{\binom{x-x_0}{k}}{\binom{x_0+k}{k}}, \qquad \Re(x) > 0.
\end{equation}

\begin{tcolorbox}
In summary, if we are given a sequence $\{a_n\}_{n\ge0}$ which arises as the values of a function whose derivatives eventually become completely monotone on $(0,\infty)$, then the Newton series in eq.~\eqref{eq:newtonseries} provides an explicit interpolation of this sequence to a function defined on the right half-plane.

\medskip
\noindent
In physics language, the Newton series plays a role analogous to a dispersion relation, reconstructing the full function from discrete data together with positivity constraints.
\end{tcolorbox}

This construction gives a constructive and analytic answer to the initial interpolation problem: not only is the function uniquely determined under suitable CM conditions, but it can also be explicitly written and analytically continued to $\Re(x) > 0$.

\subsection{Completely monotone functions in several variables}
 We will now look at the class of CM functions in several variables and on more general domains that are needed for applications in physics.

 A function $f:\mathbb{R}^n \rightarrow \mathbb{R}$ is completely monotone in $I \subseteq \mathbb{R}^n$ if it satisfies 
\begin{align}\label{cmmulti}
(-\partial_{x_1})^{m_1} \ldots (-\partial_{x_n})^{m_n} f(x_1, \ldots , x_n) \ge 0, \qquad \forall m_i\in \mathbb{Z}_{\ge0} \quad  \text{and}\quad \forall (x_1,...,x_n)\in I
\end{align}
which is just a straightforward generalization of the single variable definition in eq.\eqref{eq:def:monotonex}. 

\paragraph{Examples.}
\begin{enumerate}
\item A linear function with non-negative coefficients raised to a negative power,
\begin{align}\label{eq:linearnegativepower}
   f(x_1, \ldots , x_n) = \frac{1}{(c_1 x_1 + \ldots c_n x_n+d)^{\alpha}},\, \qquad \forall\, c_i \ge 0\,, \quad \alpha,d\ge 0.
\end{align}
This follows directly from the definition in eq.~\eqref{cmmulti}. This example though very simple is very important for many applications such as to Feynman integrals as we shall see in section~\ref{sec:feynman}.

The case of non-linear polynomials raised to negative powers is more complicated and involves hyperbolic polynomials \cite{Scott_2014,Kozhasov:2019} see the discussion around \ref{Cmhyperboly}.
\item A simple 2 variable example is 
\be
f(x_1,x_2)=\frac{1}{x_1-x_2} \log{\left( \frac{b+x_1}{b+x_2}\right)}
\ee
which is CM on $\mathbb{R}^2_{+}$ for $b\ge0$ which can be 
seen by 
 \be
f(x_1,x_2)=\int_{\mathbb{R}^2_{+}}~du~dv \frac{e^{-b (u+v)}}{u+v}\, e^{-u\, x-v\, y}
\ee
\item As a two-variable example involving polylogarithms, the 
following function appears in a finite seven-point one-loop integral \cite{Arkani-Hamed:2010zjl},
\begin{align}
\begin{split}
\Psi^{(1)}(x_1,x_2) =& \, {\rm Li}_{2}(1-x_1) +{\rm Li}_{2}(1-x_2)   + \log x_1 \log x_2 - \pi^2/6 
\,.
\end{split}
\end{align}
which is CM in the region $x_1+x_2 \le 1$. 
To see that this is CM, it is useful to consider
\begin{align}\label{defPsi}
g(x_1,x_2) =& \frac{\Psi^{(1)}(x_1,x_2)}{1-x_1 - x_2}\,,
\end{align}
which has 
the dispersive integral representation,
\begin{align}
g(x_1,x_2) =  \int_0^{\infty}\int_0^{\infty}  \tfrac{dy_1 dy_2}{(x_1 + y_1) (x_2 + y_2) (1+y_1+y_2) } \, .
\end{align}
From this equation it is manifest that $g$ is CM on $\mathbb{R}^2_{+}$.
Moreover, in view of eq. (\ref{defPsi}), we can deduce that $\Psi^{(1)}$ is CM, in the region $x_1+x_2 \le 1$.

\end{enumerate}
Multivariate completely monotone functions in projective space are characterized
by a generalization of the Bernstein--Hausdorff--Widder theorem due to
Choquet \cite{Choquet1954}. To state this result, we first clarify the notion
of complete monotonicity on cones.

\subsubsection{Completely monotone functions on projective space}

Let $V$ be a finite-dimensional vector space.\

\noindent
A \textit{cone} is a subset $C \subset V$ such that
\begin{equation}
C = { \lambda x \mid x \in C,; \lambda > 0 } ,.
\end{equation}

\medskip

The \textit{dual cone} is defined as
\begin{equation}
C^* = { y \in V^* \mid \langle y, x \rangle \ge 0 ;; \forall x \in C } ,,
\end{equation}
where $\langle y, x \rangle = \sum_i x_i y_i$.

\medskip

\begin{definition}
A function $f: C \to \mathbb{R}$ is \textit{completely monotone} (CM) if
\begin{equation}\label{def:cmcone}
(-1)^k D_{v_1} \cdots D_{v_k} f(x) \ge 0 ,,
\quad \forall v_1,\dots,v_k \in C,\; \forall x \in C,\; k = 0,1,2,\dots
\end{equation}
where $D_v$ denotes the directional derivative along $v$.
\end{definition}

\medskip

To check complete monotonicity of a function $f(\mathbf{x})$ on a convex cone
$C \subset \mathbb{R}^n$, one must in principle verify
eq.~\eqref{def:cmcone} for arbitrary directions $\mathbf{v}_i \in C$.
In practice, it is sufficient to check these derivatives along the
\textit{extremal rays} of the cone.

\medskip

\begin{definition}[Extremal ray]
A ray $R = \{ \lambda\, \mathbf{v} \mid \lambda \ge 0 \}$ is called an
\textit{extremal ray} if it cannot be written as a non-trivial sum of two
vectors in $C$. That is, if $\mathbf{v} = \mathbf{u}_1 + \mathbf{u}_2$ with
$\mathbf{u}_1, \mathbf{u}_2 \in C$, then $\mathbf{u}_1$ and $\mathbf{u}_2$
must be proportional to $\mathbf{v}$ with non-negative coefficients.
\end{definition}

\medskip

The complexity of verifying CM properties depends on whether the cone is
\textit{polyhedral} or \textit{non-polyhedral}.

\medskip

\noindent
A cone is \textit{polyhedral} if it is generated by finitely many extremal
rays ${ \mathbf{v}_1, \dots, \mathbf{v}_m }$.

\begin{itemize}
\item[1.] \textit{(The positive orthant).}
The simplest case is $C = \mathbb{R}^n_+$. The extremal rays are the
basis vectors $\mathbf{e}_i=(0,...1,...,0)$. Checking CM reduces to verifying
$(-1)^k \partial*{i_1} \cdots \partial_{i_k} f \ge 0$.

\item[2.] \textit{(A simplex cone).}
Consider $C = \{ (x,y) \mid 0 < x < y \}$, with extremal rays
$\mathbf{v}_1 = (0,1)$ and $\mathbf{v}_2 = (1,1)$.

For the function
\begin{equation}
f(x,y) = \frac{y}{x (y-x)} \,,
\end{equation}
one finds
\begin{equation}
(-1)^n D^n_{v_1} f = \frac{n!}{(y-x)^n}, \quad
(-1)^n D^n_{v_2} f = \frac{n!}{x^n}, \quad
D_{v_1} D_{v_2} f = 0 \,,
\end{equation}
showing that $f$ is CM on $C$. 

\end{itemize}

\medskip

A cone is \textit{non-polyhedral} if it requires infinitely many extremal rays.

\begin{itemize}
\item[1.] \textit{(The Lorentz/light cone).}
In $2+1$ dimensions,
\begin{equation}
C = { (t,x,y) \mid t \ge \sqrt{x^2 + y^2} }
\end{equation}
has a continuous family of extremal rays.

```
Consider
\begin{equation}
f(x,y) = \frac{1}{1 - x^2 - y^2}, \qquad x^2 + y^2 < 1 \,.
\end{equation}

This homogenizes to
\begin{equation}
\hat{f}(x,y,z) = \frac{1}{z^2 - x^2 - y^2}
\end{equation}
on the cone
\begin{equation}
C = \{ (x,y,z) \mid z \ge 0,\; z^2 - x^2 - y^2 \ge 0 \} \,.
\end{equation}

For a generic direction $v = (r\cos\theta, r\sin\theta, z)$ with $z > r$,
one finds
\begin{equation}
-D_v \hat{f} = \frac{2}{z^2 - r^2} \ge 0 \,,
\end{equation}
and more generally
\begin{equation}
(-1)^k D_{v_1} \cdots D_{v_k} \hat{f}
= \prod_i \frac{1}{z^2 - r_i^2} \ge 0 \,,
\end{equation}
showing that $f$ is CM on $C$.

\item[2.] \textit{(Positive semidefinite cone).}
The cone of symmetric positive semidefinite matrices $S^n_+$ is
non-polyhedral for $n \ge 2$. Its extremal rays correspond to rank-one
matrices $\mathbf{v}\,\mathbf{v}^T$.

\end{itemize}

\subsection{Polar Duals and Projective Duality}\label{sec:polardual}

Let $A \subset \mathbb{R}^{m+1}$ be a convex body containing the origin in its interior.
\begin{definition}(Polar dual:)
 The \textit{polar dual} of $A$ is the set
\begin{equation}
    A^* = \{ W \in (\mathbb{R}^{m+1})^* \mid W \cdot Y \ge -1 \;\; \forall\,\, Y\, \in A \}.
\end{equation}   
\end{definition}

After projectivization, one often writes
\begin{equation}
    A^* = \{ W \mid W \cdot Y \ge 0 \;\; \forall\, Y \, \in A \},
\end{equation}
depending on normalization conventions. This is the convention we adopt in the following examples.

\medskip
Geometrically, the polar dual $A^*$ is the set of hyperplanes that do not intersect the interior of $A$, and the boundary of $A^*$ corresponds to hyperplanes tangent to $A$. Duality is involutive up to closure: $(A^*)^* = A$ for sufficiently regular convex bodies.

\medskip
To construct the polar dual explicitly, we proceed in two steps:
\begin{enumerate}
    \item \textbf{Tangency condition:} A dual vector $W$ lies on the boundary of $A^*$ if it is normal to a tangent hyperplane of $A$. For a differentiable boundary, this gives
    \begin{equation}
        W = \lambda\, \nabla f(Y),
    \end{equation}
    where $f(Y) = 0$ defines the boundary of $A$.
    
    \item \textbf{Duality (normalization) condition:} The hyperplane defined by $W$ must satisfy
    \begin{equation}
        W \cdot Y = - 1,
    \end{equation}
    which fixes the scale factor $\lambda$.
\end{enumerate}

\medskip
\textbf{Examples:}
\begin{itemize}
    \item[1.] \textbf{Circle:} For the unit circle $f(x,y) = x^2 + y^2 - 1 = 0$, the tangency condition gives
    \[
        u = 2 \lambda x, \quad v = 2 \lambda y,
    \]
    and the duality condition $u\, x + v\, y = -1$ yields $\lambda = -1/2$. Substituting back, we find the dual curve
    \[
        u^2 + v^2 = 1,
    \]
    showing the circle is self-dual.
    
    \item[2.] \textbf{Cubic:} For $f(x,y) = x^3 + y^3 - 1 = 0$, the tangency condition gives
    \[
        u = 3 \lambda x^2, \quad v = 3 \lambda y^2,
    \]
    and $u\, x + v\, y = 1$ determines $\lambda$. Eliminating $(x,y,\lambda)$ gives the dual cubic
    \[
        1 - 2u^3 - 2v^3 - 2 u^3 v^3 + u^6 + v^6 = 0.
    \]

    \item[3.] \textbf{\(n\)-Sphere:} For the unit \(n\)-sphere $f(x_1,\dots,x_n) = \sum_{i=1}^n x_i^2 - 1 = 0$, tangency gives
    \[
        W = 2 \lambda (x_1,\dots,x_n),
    \]
    and the duality condition $W \cdot (x_1,\dots,x_n) = - 1$ implies $\lambda = - 1/2$. Substituting back yields
    \[
        \sum_{i=1}^n W_i^2 = 1,
    \]
    showing that the unit \(n\)-sphere is self-dual.
\end{itemize}

\medskip
 The polar dual encodes all \emph{supporting hyperplanes} of a convex body. One can think of the \textit{support function}
\[
h_A(W) = \sup_{Y \in A} W \cdot Y
\]
as measuring the maximal "shadow" of $A$ in the direction $W$. The dual $A^*$ consists of all $W$ for which this shadow is bounded by a chosen normalization.  

This construction is closely related to the \textbf{Legendre transform}:
\[
f^*(W) = \sup_Y \big( W \cdot Y - f(Y) \big),
\]
which exchanges variables for their conjugate momenta. In classical mechanics, this is exactly the step from the Lagrangian to the Hamiltonian: the dual variables $W$ play the role of momenta, arising from the tangency and duality conditions, just like velocities are replaced by momenta via a Legendre transform.



\medskip

Completely monotone functions on cones are characterized by the following
integral representation.

\begin{tcolorbox}
\begin{theorem}[\bf Bernstein--Hausdorff--Widder--Choquet]
A function $f$ on an open cone $C \subset \mathbb{R}^n$ is completely monotone
if and only if it admits the representation
\begin{equation}\label{eqn:choquet}
f(x) = \int_{C^*} e^{-\langle y, x \rangle} , d\mu(y) ,,
\end{equation}
where $\mu$ is a positive measure supported on the dual cone $C^*$.
\end{theorem}
\end{tcolorbox}

In particular if  $C = \mathbb{R}^n_+$, then $C^*=C$  and this reduces to the standard multidimensional Laplace
transform with a positive measure.

\medskip

The BHWC theorem admits a natural geometric interpretation in projective space. For example, for $\mu(y) \equiv 1$ in eq.~\eqref{eqn:choquet}:
\begin{equation}
    I(x) = \int_{C^*} e^{-\langle y, x \rangle} \, d^{n+1}y.
\end{equation}
Using the homogeneity of the cone, one can perform the radial integral:
\begin{equation}
    I(x) = \int_{C^*} \frac{1}{\langle y, x \rangle^{n}} \, d^n y,
\end{equation}
which can be interpreted as a projective integral over the dual cone.The resulting ``volume'' depends on the choice of $x$, which selects an affine slice of the dual cone, but different choices of $x$ correspond to equivalent projective measures.

\medskip

More generally, allowing a non-negative measure $\mu(y) \ge 0$ promotes this to a weighted, or ``generalized'', projective volume. 
\begin{tcolorbox}[colback=green!5!white,colframe=green!75!black]
Every completely monotone function on a convex cone can be represented as an integral over the dual cone. In projective terms, this means that each CM function secretly encodes a (possibly weighted) projective volume of the dual cone.
\end{tcolorbox}
 This as we shall see explains why such functions are natural when we talk about dual volumes and positive geometries. We will see more about this connection in section 2.

\medskip

\noindent
When a multivariate rational function is completely monotone on a convex cone, this property is intimately related to the notion of hyperbolic polynomials~\cite{Scott_2014}, \cite{Kozhasov:2019},\cite{mazzucchelli2025canonicalformsdualvolumes}. We will end this section by reviewing this connection

\subsection{Stieltjes functions in one variable}
\medskip
\noindent
Till now, we have studied functions with a \emph{real property}: complete monotonicity. While this property indirectly implies analyticity in certain strips of the complex plane, it remains quite restrictive. Many functions of interest in physics are inherently complex-valued and exhibit richer analytic structure.

\medskip
\noindent
We now turn our attention to a special subclass of completely monotone functions, known as \emph{Stieltjes functions}. These functions are particularly well-behaved: they are analytic in the cut complex plane $\mathbb{C} \setminus (-\infty,0]$ and satisfy a complex-analytic property called the \emph{Herglotz property}. Stieltjes functions obey analogues of unsubtracted dispersion relations and are remarkably amenable to rational approximation techniques, such as Padé approximations. Remarkably, using only the Taylor expansion of a Stieltjes function on the positive real axis, one can accurately approximate its values anywhere in the cut complex plane. In this sense, Stieltjes functions generalize the idea of analytically continuing a completely monotone function to the entire right half-plane via the Newton series, using only its values at the positive integers. With this motivation in mind, we now turn to a formal definition of Stieltjes functions and summarize their key properties.

\subsubsection{Defintion and basic properties}
\begin{definition}
A function $f(z)$ analytic in the cut plane $\mathbb{C}\backslash(-\infty,\, -R]$ is called a \textit{Stieltjes function} if it admits an integral representation
\begin{equation} \label{eq:Stieltjesdefinition}
f(z) = \int_{0}^{1/R} \frac{\rho(u)}{1 + u \,z}\, du \,, 
\quad \text{with} \quad 
\rho(u) \ge 0 \;\; \forall \; u \in (0, 1/R) \,.
\end{equation}
\end{definition}

\medskip

It follows from eq.~\eqref{eq:Stieltjesdefinition} that Stieltjes functions form a subset of completely monotonic (CM) functions, since
\begin{align}\label{eq:muforStieltjes}
\left( - \partial_z \right)^{n} f(z) 
= \int_{0}^{1/R} du\, \frac{u^{n}\, n!\, \rho(u)}{(1 + u z)^{n+1}} \ge 0 \,.
\end{align}

\noindent
\textbf{Remark.} In the literature, one also commonly defines a Stieltjes function by the representation
\begin{equation}
f(z) = \int_{0}^{1/R} \frac{\rho(u)}{z + u}\, du \,.    
\end{equation}
The form \eqref{eq:Stieltjesdefinition} used here is more natural for Padé approximation applications. The two representations are equivalent and can be mapped into each other by a simple change of variables.

\subsection*{Examples}

\begin{itemize}
\item[1.]  Rational example:
\begin{equation}
f(z) = \frac{1}{1+z} = \int_0^1 du\, \delta(u-1)\frac{1}{1+uz}.
\end{equation} 
\item[2.] Logarithmic example:
\begin{equation}
 f(z) = \frac{\log(1+z)}{z} = \int_0^1 \frac{du}{1 + u z}.   
\end{equation}

\item[3.] Polylogarithms:
\begin{equation}
\frac{\operatorname{Li}_{n+1}(-z)}{-z} = \int_0^1 dt\, \frac{(-\log \,t)^n}{1+z\, t}.
\end{equation}
\end{itemize}

\noindent In principle, the measure $\rho(u)$ can be recovered from $f(z)$ via the inversion formula
\begin{equation}
\rho(u) = \frac{\text{Disc}\, f(-1/u)}{-u}, \qquad 
\text{Disc}\, f(x) = \lim_{\epsilon \to 0} (f(x+i\epsilon) - f(x-i\epsilon))\,,
\end{equation}
but in practice this inversion is cumbersome. This motivates alternative characterizations.
Widder provided a real-variable derivative criterion similar to eq.~\eqref{eq:def:monotonex} for completely monotone functions,
\begin{theorem}( Widder's real variable chareterization)
A real-valued function defined on $(0,\,\infty)$ is Stieltjes if and only if
\begin{equation}\label{eq:WidderStieltjes}
(-1)^n (x^k f(x))^{(n+k)} \ge 0 \qquad \forall n,k\ge 0, \ x>0.
\end{equation}
\end{theorem}

When $k=0$, this reduces to the conditions defining CM functions. Stieltjes functions require additional positivity conditions on derivatives, which ensures that they can be analytically continued to the full cut plane $\mathbb{C}\setminus(-\infty,0]$, whereas CM functions generally extend only to the right half-plane.

\medskip
\noindent
 We have eq.~\eqref{eq:Stieltjesdefinition}  which is like the Laplace transform characterization of completely monotone functions  on $(0,\,\infty)$ namely eq.~\eqref{eq:BHW} . And we have Widder's condition eq.~\eqref{eq:WidderStieltjes} on the derivatives being positive analogous to eq.\eqref{eq:def:monotonex} for CM functions.
 
\medskip  

While these conditions are elegant, they are not always practical for verification. More convenient characterizations are complex-analytic in nature.

\medskip 

Following ref.~\cite[Sec.~8.6]{Bender}, it is sufficient to verify the following properties to prove a function is Stieltjes:

\begin{enumerate}
    \item \textit{Analyticity:} $f(z)$ is analytic in the cut complex plane $\mathbb{C}\setminus(-\infty,-R]$.
    \item \textit{Asymptotic behavior:} $f(z) \to C \ge 0$ as $|z|\to \infty$.
    \item \textit{Herglotz property:} $-f(z)$ is Herglotz, i.e.
    \begin{align}
        \Im f(z)\, \Im z < 0 \quad \forall\, z\notin \mathbb{R}.
    \end{align}
\end{enumerate}

\medskip
\noindent
These conditions imply the Stieltjes representation~\eqref{eq:Stieltjesdefinition} as follows. Consider the function
\[
\frac{f(w)-C}{w},
\]
where the subtraction ensures the integrand decays as $|w|^{-2}$ at infinity. Applying Cauchy's theorem on a keyhole contour around the cut $(-\infty,-R]$ we get
\[
f(z)-C = \frac{1}{2\pi i} \int_{-\infty}^{-R} \frac{\text{Disc}\, f(x)}{x-z} \, dx, \qquad 
\text{Disc}\, f(x) = \lim_{\epsilon\to0} (f(x+i\epsilon)-f(x-i\epsilon)).
\]
The Herglotz property guarantees $\Im f(x+i0)\le 0$, allowing us to define the positive measure
\[
\rho(u) = -\frac{1}{\pi \,  u} \Im f\Big(-\frac{1}{u}+i0\Big), \quad u\in(0,\, 1/R),
\]
and changing variables $x=-1/u$ yields
\[
f(z) = C + \int_0^{1/R} \frac{\rho(u)}{1+u\, z}\, du,
\]
which is exactly the Stieltjes representation. The constant $C$ can be absorbed into the measure if desired.

\medskip
\noindent
So in summary condition 1 allows us to apply Cauchy's theorem in the cut plane, condition 2 tells us we need at most one  subtractions to make the function decay for large argument and condition 3 ensures that the discontinuity across the branch cut is positive and together they imply

\begin{tcolorbox}[colback=green!5!white,colframe=green!75!black,title=Stieltjes functions as subtracted dispersion relations]
A Stieltjes function is precisely a function that admits a once-subtracted dispersion relation with a positive discontinuity across the branch cut.

\medskip

This observation explains why such functions naturally arise in many physical applications.
\end{tcolorbox}

\medskip

\noindent An alternative version is due to Krein's
\begin{theorem}[Krein's characterization]
A function $f:(0,\infty)\to\mathbb{R}$ is Stieltjes if and only if $f(x)\ge 0$ for $x>0$, and $f$ admits an analytic continuation to $\mathbb{C}\setminus(-\infty,0]$ such that $\Im f(z) \le 0$ whenever $\Im z \ge 0$.
\end{theorem}

Thus, these conditions provide a direct analytic route to proving that a function is Stieltjes, without the need for inverting measures explicitly or checking infinitely many inequalities are satisfied. 

\medskip 

The key ingredient in these characterizations is the Herglotz property, whose consequences we now briefly discuss.
The Herglotz property 
\begin{equation}\label{eq:Herglotz}
\Im f(z)\, \Im z < 0 \quad \forall\, z \notin \mathbb{R}
\end{equation}
means that $f$ maps the upper and lower half-planes to opposite half-planes. Despite the apparent simplicity of this condition, it imposes strong constraints on the analytic structure of $f$. In particular:

\begin{itemize}
    \item[1.] All poles and zeros of $f$ lie on the real axis.
    \item[2.] All poles and zeros are simple, with poles having positive residues.
\end{itemize}

\noindent
To see this, consider the local behavior of $f(z)$ near a singularity or zero $z=z_0$:
\begin{equation}
  f(z) \sim A\, (z-z_0)^{\gamma},
\end{equation}
where $\gamma \in \mathbb{R}$. Writing $(z-z_0)= r e^{i\theta}$ and $A = A_0 e^{i\phi}$ with $A_0,r>0$, we obtain
\begin{equation}
  \Im f(z) \sim A_0\, r^\gamma \sin(\gamma \theta + \phi).
\end{equation}
For $z$ approaching $z_0$ from the upper half-plane, $\theta \in (0,\pi)$. The Herglotz property requires $\Im f(z) < 0$ throughout this interval, so the function $\sin(\gamma\theta+\phi)$ cannot change sign. This is only possible if the total variation of the argument is at most $\pi$, which implies
\[
|\gamma| \le 1.
\]
For meromorphic functions, $\gamma \in \mathbb{Z}\setminus\{0\}$, hence $\gamma = \pm 1$, showing that all poles and zeros are simple.

\medskip
\noindent
For a pole ($\gamma=-1$), we require $\sin(\phi-\theta)<0$ for all $\theta\in(0,\pi)$. Taking limits:
\begin{itemize}
    \item $\theta \to 0^+$ gives $\sin(\phi)\le 0$,
    \item $\theta \to \pi^-$ gives $\sin(\phi-\pi)\le 0$, i.e. $\sin(\phi)\ge 0$.
\end{itemize}
Thus $\sin(\phi)=0$, and the inequality fixes $\phi=0$, so $A=A_0>0$. Hence all poles have positive real residues.

\medskip
\noindent
For a zero ($\gamma=1$), the condition $\sin(\theta+\phi)<0$ for all $\theta\in(0,\pi)$ forces $\phi=\pi$, so $A=-A_0$.

\medskip

\noindent
\textbf{Summary.} To verify that a function is Stieltjes, it suffices in practice to check that:
\begin{itemize}
    \item $f(x)\ge 0$ for $x>0$,
    \item $f$ satisfies the Herglotz property.
\end{itemize}

\medskip
\noindent
These constraints immediately exclude certain completely monotone functions. For example,
\[
f(x)=\frac{1}{(1+x)^2}
\]
is completely monotone but not Stieltjes, as it has a double pole, which is forbidden. Similarly,
\[
f(x)=\frac{1}{x^3+x}
\]
is completely monotone on $(0,\infty)$ but not Stieltjes, since it has poles off the real axis.
We end this discussion by pointing out that unlike CM functions Stieltjes functions are not closed under derivatives or products as the above examples clearly show. The main reason for this is the tension with the Herglotz property. Nonetheless we now metion a related class of functions which one would get if we take derivatives of a Stieltjes function. 
\subsubsection{Generalized Stieltjes Functions}

A natural extension of the space of Stieltjes functions, denoted $\mathcal{S}_{\lambda}$, is defined by the following integral representation (cf.\ ref.~\cite{KarpPrilepkina2012}):
\begin{equation}\label{eq:GSF_def}
f(z) = \int_{0}^{1/R} \frac{\rho(u)}{(1 + u\, z)^{\lambda}} \, du, \quad \text{with} \quad \rho(u) \ge 0, \ \lambda > 0.
\end{equation}
When $\lambda = 1$, we recover the standard Stieltjes class.

\subsubsection*{Key Properties}

\begin{itemize}
    \item \textbf{Geometric Mapping (The Wedge Property):} 
    While a standard Stieltjes function maps the upper half-plane (UHP) to the lower half-plane (LHP), a GSF of order $\lambda$ maps the UHP to a specific \textit{wedge}. For $z \in \text{UHP}$, the phase of the function is constrained by:
    \begin{equation}
    -\min(\pi, \pi\lambda) < \arg f(z) < 0.
    \end{equation}
    For $\lambda < 1$, the image is a narrow wedge of angle $\pi\lambda$ within the LHP. For $\lambda > 1$, the image covers the entire LHP and can extend onto further Riemann sheets.

    \item \textbf{ (Nested Inclusivity):} 
    The classes $\mathcal{S}_{\lambda}$ satisfy a nested inclusion property. If $\alpha < \beta$, then:
    \begin{equation}
    \mathcal{S}_{\alpha} \subset \mathcal{S}_{\beta}.
    \end{equation}
    This implies that any standard Stieltjes function ($\lambda=1$) is automatically a generalized Stieltjes function of order $\lambda$ for all $\lambda > 1$.

    \item \textbf{Conversion to Standard Stieltjes Form:} 
    A GSF of order $\lambda > 1$ can be mapped back to the standard $\mathcal{S}_1$ class through a power transformation of the coordinate. Specifically, if $f(z) \in \mathcal{S}_{\lambda}$, then:
    \begin{equation}
    g(z) = f(z^{1/\lambda}) \implies g(z) \in \mathcal{S}_1.
    \end{equation}
\end{itemize}

Like we alluded to earlier product of Stieltjes functions is not Stieltjes, we now list what properties preserve the property.
\subsubsection{Closure Properties}
The definition of Stieltjes functions through the Herglotz mapping and the intgeral representation leads to a remarkably rigid set of algebraic properties. Many of these follow from Krein's characterization \cite{85ec3dd0db3311dd9473000ea68e967b}:

Let $f(z), g(z)$ be Stieltjes functions. Then:

\begin{enumerate}
\item {\bf Convex cone:} For any $\lambda, \mu > 0$, $\lambda f(x) + \mu g(x)$ is Stieltjes.
\begin{quote}
\textit{Why:} The sum of two measures $d\mu_1 + d\mu_2$ remains a positive measure, and the LHP-mapping property is preserved under linear combinations with positive coefficients.
\end{quote}

\item {\bf Closure under Powers:} For $0 \leq \alpha \leq 1$, $h(x) = f(x)^\alpha$ is Stieltjes. 
\begin{quote}
\textit{Why:} If $\arg f \in (-\pi, 0)$, then $\arg f^\alpha \in (-\pi\alpha, 0)$. For $\alpha \le 1$, this range is contained within $(-\pi, 0)$, maintaining the Anti-Herglotz property.
\end{quote}

\item {\bf Closure under Hölder products:} For $0 \leq \alpha \leq 1$, $h(x) = f(x)^\alpha g(x)^{1-\alpha}$ is Stieltjes. 
\begin{quote}
\textit{Why:} The phase of the product is the convex combination of the phases: $\arg h = \alpha \arg f + (1-\alpha) \arg g$. Since both phases are in $(-\pi, 0)$, their weighted average is also in $(-\pi, 0)$.
\end{quote}

\item {\bf Inversion and Scaling:}
\begin{itemize}
    \item $h(x) = \frac{1}{f(1/x)}$ is Stieltjes.
    \item $h(x) = \frac{1}{x f(x)}$ is Stieltjes.
    \item For $\lambda > 0$, $\frac{f(x)}{\lambda f(x) + 1}$ is Stieltjes.
\end{itemize}
\begin{quote}
\textit{Why:} These transformations effectively swap the LHP and UHP or invert the variable $z$. For example, $1/z$ maps the UHP to the LHP, and the homographic transform is a conformal map that preserves the half-plane geometry.
\end{quote}

\item {\bf Composition:} $f \circ \frac{1}{g}$ and $\frac{1}{f \circ g}$ are Stieltjes.
\begin{quote}
\textit{Why:} The inner function $1/g$ maps the UHP to the UHP; the outer function $f$ then maps that UHP to the LHP, satisfying the Stieltjes condition.
\end{quote}

\item {\bf Closure under limits:} If $\{f_n(x)\}$ is a sequence of Stieltjes functions converging pointwise, $\lim_{n \to \infty} f_n(x)$ is Stieltjes. 
\begin{quote}
\textit{Why:} The set of positive measures is closed under weak convergence, ensuring the limit maintains a valid spectral representation.
\end{quote}

\item {\bf Möbius Transformations:}
$g(z) = \frac{1}{cz+d} f\left( \frac{az+b}{cz+d} \right)$ for $a,b,c,d \ge 0$ and $ad-bc > 0$ is Stieltjes.

\begin{quote}
Substituting the integral representation $f(z) = \int \frac{\mu(t)}{1+zt} dt$ into $g(z)$ yields:
\begin{equation}
g(u) = \int_{0}^{1/R} \frac{\mu(t)}{(bt+d)\left( 1 + u \frac{at+c}{bt+d} \right)} \, dt.
\end{equation}
By defining the change of variables $y = \frac{at+c}{bt+d}$, we get:
\begin{equation}
g(u) = \int_{c/d}^{\frac{a+cR}{b+dR}} \frac{1}{1+u\, y} \underbrace{\left[ \frac{\mu(t(y))}{a-b\,y} \right]}_{\nu(y)} dy.
\end{equation}
Since $a,b,c,d \ge 0$ and $ad-bc > 0$, the mapping $t \mapsto y$ is monotonic and $a-by > 0$ on the support. Thus, $\nu(y) \ge 0$, and $g(u)$ is confirmed to be a Stieltjes function.
\end{quote}
\end{enumerate}
Note that the fractional linear transformation $g(z) = \frac{1}{cz+d} f(\frac{az+b}{cz+d})$ is distinct from the composition property $f \circ (1/g)$. While composition involves nesting two Stieltjes functions, the FLT represents a symmetry of the Stieltjes class itself. The pre-factor $(cz+d)^{-1}$ is essential.

\medskip

\noindent
\medskip
\noindent
If we expand the integral representation eq.~\eqref{eq:Stieltjesdefinition} around $z \to 0$ we get
\begin{align} \label{asyexp}
f(z) \sim \sum_{n=0}^{\infty} (-1)^n a_n z^n, \qquad a_n > 0,
\end{align}
where the coefficients are given by the moments of the measure,
\begin{equation}\label{moments}
a_n = \int_0^{1/R} u^n \rho(u)\, du.
\end{equation}

\noindent
When $R>0$, the integration range is finite and all moments exist. However, for $R=0$ the integral extends to infinity, and the moments need not exist. In this case, the expansion around $z=0$ may involve logarithmic terms rather than a pure power series.

\medskip
\noindent
In practice, this issue can be avoided by expanding around any point $z_0>0$. Indeed, the shifted function
\begin{equation}
g(z) = f(z+z_0)
\end{equation}
is again Stieltjes by property 7 with $a=1,\, b=z_0,\, c=0,\, d=1$ , and takes  $R \mapsto R+z_0 >0$. This ensure the existence of a moment expansion of the form eq.~\eqref{asyexp}

\medskip
\noindent
\textbf{Example.} The function
\begin{equation}
f(z) = \frac{\log z}{z-1}
=
\int_0^\infty \frac{du}{(1+z\,u)(1+u)}
\end{equation}
does not admit finite moments at $z=0$. However,
\begin{equation}
g(z) = f(1+z) = \frac{\log(1+z)}{z}
=
\int_0^1 \frac{du}{1+zu}
\end{equation}
does. 

\medskip

\noindent
The existence of moment expansions, together with the positivity and controlled analytic structure of Stieltjes functions, makes them particularly well-suited for rational approximation. These features lead to powerful convergence results for Padé approximants, to which we now turn.

\subsection{Padé approximation and convergence theorems}
\label{sec:subsecPadeconvergence}

The Padé approximation (PA) is a simple and useful alternative to polynomial approximations of analytic functions. Suppose we are given the Taylor expansion of a function $f(z)$ about a point $x_0$,
\begin{equation}
f(z) = \sum_{k=0}^{\infty} a_k (z-x_0)^k,
\end{equation}
which converges in some neighborhood of $x_0$.

\medskip
\noindent
To construct the Padé approximant, one proceeds as follows:
\begin{itemize}
    \item Truncate the Taylor series to $K=N+M+1$ terms.
    \item Find a rational function
    \begin{equation}
    P^N_M(z;x_0) = \frac{A_0 + A_1 (z-x_0) + \cdots + A_N (z-x_0)^N}
    {1 + B_1 (z-x_0) + \cdots + B_M (z-x_0)^M},
    \end{equation}
    such that its Taylor expansion agrees with that of $f(z)$ up to order $N+M+1$.
\end{itemize}

\medskip

Thus, the Padé approximant provides a rational approximation to the analytic function $f(z)$.

\medskip
\noindent
It has two key advantages:
\begin{itemize}
    \item \textit{Analytic continuation:} It allows one to evaluate the function outside the radius of convergence of the original Taylor series.
    \item \textit{Series acceleration:} It often converges much faster than the Taylor series within its domain of convergence.
\end{itemize}
\noindent
Padé approximations can be applied to any function but in particular the mathematical theory is very well developed and understood for Stieltjes functions. In particular,
for Stieltjes functions analytic in $\mathbb{C}\setminus(-\infty,-R]$, Padé approximants satisfy the following remarkable properties:

\begin{itemize}

\item[(a)] \textit{Convergence on the real axis.}  
For $x \in \mathbb{R}\setminus(-\infty,-R]$ with $x \ge x_0$:
\begin{itemize}
\item The sequence $\{P^N_N(x;x_0)\}$ is monotonically decreasing in $N$.
\item The sequence $\{P^{N-1}_N(x;x_0)\}$ is monotonically increasing in $N$.
\item These sequences bound the function:
\begin{equation}
P^{N-1}_N(x;x_0) \le f(x) \le P^N_N(x;x_0), \qquad x \ge x_0.
\end{equation}
\end{itemize}

\item[(b)] \textit{Poles and residues.}  
For $J \ge -1$, the Padé approximants $P^{N+J}_N(z;x_0)$ have only simple poles located on $(-\infty,-R)$ with positive residues. In particular,
\begin{equation}
P^{N-1}_N(z;x_0) = \sum_{i=1}^N \frac{\beta_i}{1+\gamma_i z}, \qquad \beta_i,\gamma_i \ge 0.
\end{equation}

\item[(c)] \textit{Convergence in the cut plane.}  
In $\mathbb{C}\setminus(-\infty,-R]$, the sequences $\{P^N_N(z;x_0)\}$ and $\{P^{N-1}_N(z;x_0)\}$ converge to $f(z)$, provided the coefficients of the asymptotic expansion satisfy
\begin{equation}
|a_n| = O((2n)! \, C^n)
\end{equation}
for some constant $C$.

\item[(d)] \textit{Error bounds.}  
Let $\Delta>0$ and define
\begin{equation}
\mathcal{D}^+(\Delta) = \{x+iy \in \mathbb{C} \mid x \le -R, |y| \ge \Delta\} \cup \{x+iy \in \mathbb{C} \mid x>-R\}.
\end{equation}
If the asymptotic expansion of $f(z)$ is convergent, then for $z \in \mathcal{D}^+(\Delta)$,
\begin{align}
|f(z)-P^{M+J}_M(z;x_0)| 
< c \left|\frac{z-x_0}{\rho}\right|^{J+1}
\left| \frac{\sqrt{\rho+z-x_0}-\sqrt{\rho}}{\sqrt{\rho+z-x_0}+\sqrt{\rho}} \right|^{2M},
\end{align}
for all $J \ge -1$, $M \ge 1$, where $\rho = R + x_0 - \Delta$ and $c$ is a constant.

\end{itemize}

\medskip
\noindent
Thus, Padé approximants provide convergent rational approximations to Stieltjes functions on the real axis and in the cut complex plane, with controlled analytic structure and rigorous error bounds.

\subsubsection{Multivariate Stieltjes functions}

Similar to univariate case, inthe multivariate case also there are two closely related integral representations of Stieltjes functions,though these seem to not be obviously equivalent. Let $f(z_1,\dots,z_n)$ be analytic in $(\mathbb{C}\backslash(-\infty,0])^n$. Then $f$ is considered a multivariate Stieltjes function if it admits either of the following forms:

\begin{itemize}
\item \textbf{Version 1 (CM functions with a CM measure perspective):}
\begin{equation}
f(z_1,\dots,z_n)
= \int_0^{\infty}\!\dots\!\int_0^{\infty} dt_1\dots dt_n\,
\frac{\mu(t_1,\dots,t_n)}{(t_1+z_1)\dots(t_n+z_n)},
\end{equation}
with $\mu(t_1,\dots,t_n)\ge0$. This is a direct generalization of univariate CM/Stieltjes functions to multiple variables. Each variable $z_i$ enters via a separate CM kernel.

\item \textbf{Version 2 (multivariate Padé perspective):}
\begin{equation}
f(z_1,\dots,z_n)
= \int_0^{\infty}\!\dots\!\int_0^{\infty} dt_1\dots dt_n\,
\frac{\mu(t_1,\dots,t_n)}{1+t_1 z_1+\dots+t_n z_n}.
\end{equation}
Version 2 is particularly natural when constructing multivariate Padé approximants, as the denominator is linear in each variable, which preserves monotonicity and positivity properties necessary for convergence results analogous to the univariate case~\cite{Barnsley:1978}.
\end{itemize}

A concrete example arises in $\mathcal{N}=4$ SYM theory. Consider the six-particle MHV remainder/ratio function, or equivalently, the four-point in-in correlator of a conformally coupled scalar exchange on a de Sitter background:
\begin{equation}
f(u,v)=\frac{Li_2(1-u)+Li_2(1-v)+\log u\, \log v-\zeta_2}{1-u-v}.
\end{equation}
This function can be written in both Version 1 and Version 2 forms:

\begin{align}
\text{Version 1:} \quad & f(u,v) = \int_{\mathbb{R}^2_+} dx\,dy\, \frac{1}{(x+u)(y+v)(1+x+y)},\\
\text{Version 2:} \quad & f(u,v) = \int_{\mathbb{R}^2_+} dx\,dy\, \frac{1}{(1+x)(1+y)(1+x u+y v)},
\end{align}
where the second representation is obtained by a change of variables $x\rightarrow x u$, $y\rightarrow y u$.

However, not all multivariate functions are Stieltjes in both representations. For instance, the Mandelstam representation of the one-loop Coulomb branch amplitude in $\mathcal{N}=4$ SYM~\cite{Caron-Huot:2014lda},
\begin{equation}
f_1(u,v)=\int_{0}^{1} dx \int_{0}^{1-x} dy\,
\frac{1}{(x+u)(y+v)\sqrt{1-x-y}},
\end{equation}
fits manifestly only into Version 1. Understanding which representation is appropriate in general remains an open question for more complicated multivariate integrals.






\newpage
\section{Positivity properties of amplitudes and related observables}

We now review several examples of building blocks like Feynman integrals and observables such as amplitudes in quantum field theory that exhibit the positivity properties discussed above. These examples were identified in ref.~\cite{henn2025positivitypropertiesscatteringamplitudes}, and we limit ourselves here to a brief review of a few of them, referring the reader to the original reference for other examples and more details.

\medskip

Broadly speaking, the origin of these positivity properties can be traced to three distinct mechanisms:
\begin{enumerate}
    \item \textbf{Representation-theoretic arguments:} Positivity follows from explicit integral representations with manifestly positive kernels or can be argued for directly from the explicit result.
    
    \item \textbf{Analyticity and causality:} General properties of the S-matrix, such as analyticity, unitarity, and crossing symmetry, lead to dispersion relations and in-turn positivity.
    
    \item \textbf{Positive geometry:} In $\mathcal{N}=4$ SYM, observables admit geometric interpretations and positivity is expected from such considerations.
\end{enumerate}

\subsection{Completely monotone structure of the cusp anomalous dimension} \label{CAD}

A prominent example where complete monotonicity arises in quantum field theory is the \emph{angle-dependent cusp anomalous dimension}. Consider a Wilson line containing a cusp with opening angle $\phi$. Its expectation value exhibits an ultraviolet divergence of the form
\begin{equation}
\langle W_{\text{cusp}} \rangle \sim \frac{1}{\epsilon}\, \Gamma_{\text{cusp}}(\phi),
\end{equation}
where $\Gamma_{\text{cusp}}(\phi)$ governs infrared singularities of scattering amplitudes \cite{Korchemsky:1987wg}, anomalous dimensions of large-spin operators \cite{Polyakov:1980ca, Korchemsky:1988si}, and the quark--antiquark potential \cite{Drukker:1999zq}.

If variable $x = e^{i\phi}$ and we restrict to the Euclidean region $x \in (0,1)$ then the cusp anomalous dimension admits a perturbative expansion of the form
\begin{equation}
\Gamma_{\text{cusp}}(x) = \sum_{L \ge 1} g^{2L}\, \Gamma_{\text{cusp}}^{(L)}(x).
\end{equation}

\begin{figure}[H]
    \centering
\includegraphics[width=0.5\linewidth]{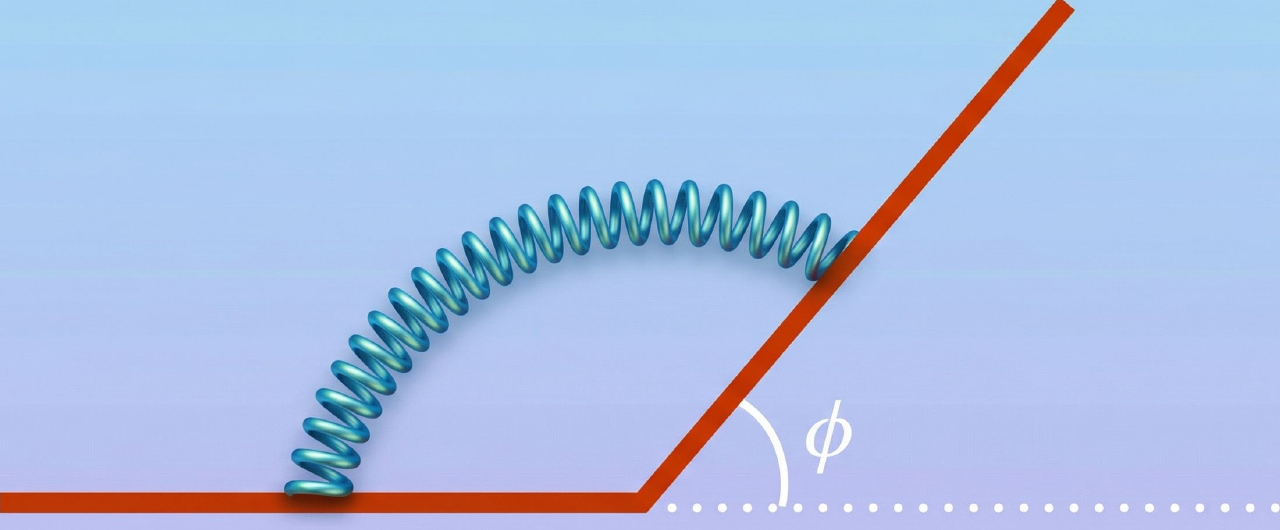}
    \caption{Wilson line containing a cusp with opening angle $\phi$.}
    \label{fig:cusppic}
\end{figure}
\medskip
The result from ref.~\cite{henn2025positivitypropertiesscatteringamplitudes} is:
\begin{tcolorbox}[colback=blue!5!white,colframe=blue!60!black,title=\textbf{Complete monotonicity of the angle dependent cusp anomalous dimension.}]

The \emph{angle-dependent cusp anomalous dimension}
\begin{equation}
(-1)^{L+1} \, \Gamma_{\text{cusp}}^{(L)}(x)
\end{equation}
is a completely monotone function in $x \in (0,1)$. 

This has been checked based on available perturbative data up to $L=3$ in QCD and $\mathcal{N}=4$ SYM, and up to $L=4$ in QED.
\medskip
\end{tcolorbox}

\medskip

As a simple illustration, the one-loop result is given by
\begin{equation}\label{cusp1loop}
\Gamma^{(1)}_{\rm cusp}(x) = \frac{1-x}{1+x} (-\log x),
\end{equation}
which is manifestly completely monotone on $(0,1)$, since all 3 factors $1-x$, $\frac{1}{1+x}$ and $-\log{x}$ are all CM on $(0,1)$ and the property is preserved when we take products.

\medskip

Has been checked in planar $\mathcal{N}=4$ SYM upto 3 loops \cite{Henn:2013wfa} but also QCD at three loops \cite{Grozin:2014hna} and in QED at four loops \cite{Bruser:2020bsh}.


In the Euclidean region $x \in (0,1)$, the cusp anomalous dimension diverges logarithmically as $x \to 0$,
\begin{equation}
\Gamma_{\rm cusp}(x) \sim - \Gamma^{\infty}_{\rm cusp} \log x,
\end{equation}
while it vanishes at $x=1$. If one rescales the functions to match their small-$x$ asymptotics, their shapes become remarkably similar across different theories (see Fig.~\ref{fig:CM_Plot}), despite differing numerically at the level of a few percent.

\medskip

\subsection{Scalar Feynman Integrals}\label{sec:feynman}
Scalar Feynman integrals are the basic building blocks of the  amplitudes in QFT. They are given by Feynman parametrization (see ref~\cite{Weinzierl:2022eaz}) for a given Feynman graph $G$ with $L$ loops in $D$ dimensions and take the form
\begin{equation} \label{eq:FeynmanInt}
I(x_i)
=
\frac{
\Gamma\!\left(\sum_i \nu_i - \tfrac{L D}{2}\right)
}{
\prod_i \Gamma(\nu_i)
}
\int_{\alpha_i \ge 0}
\prod_i d\alpha_i \; \delta\bigg(1-\sum_i\alpha_i\bigg) 
\frac{
 \, \alpha_i^{\nu_i - 1}
\;
U(\alpha)^{\sum_i \nu_i - (L+1)\tfrac{D}{2}}
}{
 \, F(\alpha_i,x_i)^{\sum_i \nu_i - \tfrac{L D}{2}}
}
\; .
\end{equation}
Where $U$ and $F$ are Graph polynomials defined by
\begin{equation}
U(\alpha)
=
\sum_{T \in \mathcal{T}_1}
\prod_{e_i \notin T} \alpha_i \, ,
\end{equation}
\begin{equation}
F(\{\alpha_i\},\{x_i\})
=
\sum_{(T,R) \in \mathcal{T}_2}
\left(
\prod_{e_i \notin (T,R)} \alpha_i
\right)
\left(
- s_{T,R}
\right)
+
U(\alpha)
\sum_{i=1}^{n} \alpha_i m_i^2 \, .
\end{equation}
where $T_1$ and $T_2$ are spanning trees and two-forests respectively. The kinematical variables are given by 
\begin{align}
    s_{(T,R)} &= \bigg(\sum_{e_i\notin(T,R)}q_i\bigg)^2\\
    \{x_i\}&=\{-s_{(T;R)},m_i^2\}.
\end{align}

\noindent The Euclidean region is defined as 
\begin{equation} \label{eq:Euclideandef}
\mathcal{E}= \{\{x_i\}~|~ F(\{\alpha_i\};\{x_i\}) \ge 0~~ \forall~~\alpha_i \ge 0 \}  \,.  
\end{equation}
For example, for the massive bubble integral, we have 
\begin{equation}
\begin{aligned}\label{eq:bubba1}
 F(\alpha_1,\alpha_2;s,m^2) =& \alpha_1\alpha_2 (-s)+ m^2 (\alpha_1+\alpha_2)^2
 \end{aligned}
\end{equation}
and demanding eq.\eqref{eq:Euclideandef} gives: 
\begin{align}\label{eq:bubbleLandau}
 \mathcal{E}(s,m)&= \{(s,m)~|~ s \le 4\,m^2 \}  \,.
\end{align}  

Another example is the massive box integral with equal internal mass. Here we have 
\begin{equation}
\begin{aligned}\label{eq:massbox}
F(\alpha_1,...,\alpha_4;s,t,m^2) =& \alpha_1 \alpha_3 (-s)+   \alpha_2 \alpha_4 (-t)+(\alpha_1+\alpha_2+\alpha_3+\alpha_4)^2~m^2
\end{aligned}
\end{equation}
and eq.~\eqref{eq:Euclideandef} gives
\begin{eqnarray}\label{eqn:massbocLandau}
\mathcal{E}(s,t,m^2)&=&\{(s,t,m)~|~s\leq 4m^2,~t \leq 4m^2\} \,.  
\end{eqnarray}
The Euclidean region admits a natural geometric interpretation as a \emph{copositive cone}, see ref.~\cite{sturmfels2025copositivegeometryfeynmanintegrals}. 
Within this region, scalar Feynman integrals exhibit remarkable positivity properties. 
In ref.~\cite{henn2025positivitypropertiesscatteringamplitudes}, it was shown that scalar Feynman integrals are completely monotone (CM), and subsequently in ref.~\cite{ditsch2026approximatingfeynmanintegralsusing} that, under suitable conditions, they are in fact Stieltjes functions.

The origin of these properties can be traced back to the Feynman parameter representation in eq.~\eqref{eq:FeynmanInt}. 
There, the dependence on the kinematic variables $x_i$ enters exclusively through the second Symanzik polynomial $F(x_i,\alpha_i)$, which is linear in the $x_i$ and appears raised to a negative power. 
Since functions of this type are completely monotone (and, under stronger conditions, Stieltjes), it follows that the full Feynman integral inherits these properties, provided the integral converges. 
Indeed, positive linear combinations preserve both complete monotonicity and the Stieltjes property.

There are, however, two important subtleties:
\begin{enumerate}
\item The kinematic variables $x_i$ are not all independent, and one must therefore choose a set of independent variables before establishing complete monotonicity.
\item For special kinematic configurations, the Euclidean region itself may be empty, obstructing the applicability of these arguments.
\end{enumerate}

These issues do not arise for planar integrals, but they can occur in the non-planar case. 
In such situations, an additional technical condition on the second Symanzik polynomial is required.

A simple example illustrating these subtleties is provided by massless four-particle scattering. 
In this case, the Mandelstam variables $s,t,u$ satisfy $s+t+u=0$, so that the variables $y=\{-s,-t,-u\}$ cannot all be positive simultaneously. 
As a result, the Euclidean region is empty, and the second Symanzik polynomial need not be positive definite, as happens for instance in the non-planar double box integral. 
Moreover, all three variables enter non-trivially in $F$, making it impossible to isolate an independent positive set.

This situation changes if one of the external legs is taken off-shell. 
For example, if $p_4^2 \neq 0$, then the relation becomes $s+t+u=p_4^2$, and the Euclidean region is non-empty. 
In this case, one can trade $u$ for $s,t,p_4^2$, which form an independent set of variables, allowing the previous arguments to go through.
\begin{figure}[H]
    \centering
    \includegraphics[width=0.9\linewidth]{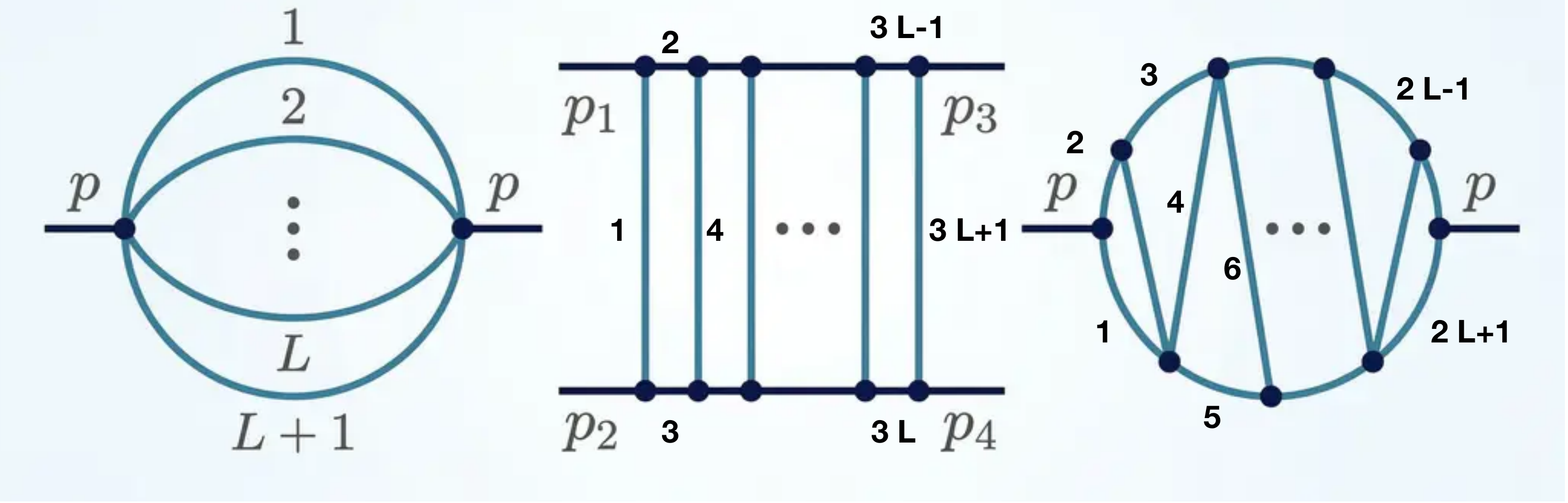}
    \caption{Some families of Feynman integrals that are Stieltjes (a) Banana integrals in $D=2$ (b) box intgerals in $D=6$ (c) zig-zag integrals in $D=4$. }
    \label{fig:placeholder}
\end{figure}

\medskip
We can now summarize the results of refs.~\cite{henn2025positivitypropertiesscatteringamplitudes,ditsch2026approximatingfeynmanintegralsusing}.
\begin{tcolorbox}[colback=blue!5!white,colframe=blue!60!black,title=Sufficient condition for scalar Feynman integrals to be completely monotone functions,]
\begin{theorem}[ see~\cite{henn2025positivitypropertiesscatteringamplitudes},\cite{ditsch2026approximatingfeynmanintegralsusing}]
Scalar Feynman integrals, as defined in eq.~(\ref{eq:FeynmanInt}), are completely monotone functions in the variables $x_i$ in the Euclidean region if 
the second Symanzik polynomial can be written in the form

\medskip

\begin{equation}
F = \sum_i\,A_i\, x_i,
\end{equation}
with $A_i \ge 0$.

\medskip

For planar Feynman integrals this condition is satisfied generally and we can choose $\{x_i\}=\{-s_{T,R}, m_i^2\}$.
For non-planar integrals, there is strong evidence from explicit examples that the above condition on the Symanzik polynomial continues to hold, although a general proof remains an open problem.
\end{theorem}
\end{tcolorbox}
\medskip

We have used the multi-variate version of complete monotonicity eq.\eqref{cmmulti}. However for the Stieltjes we stick to the single variate version with the understanding that all other variables are held fixed in the Euclidean region.
\begin{tcolorbox}[colback=blue!5!white,colframe=blue!60!black, title=Sufficient condition for scalar Feynman integrals to be Stieltjes functions]
\begin{theorem}[see ref.~\cite{ditsch2026approximatingfeynmanintegralsusing}]
Scalar Feynman integrals, as defined in eq.~(\ref{eq:FeynmanInt}), are Stieltjes functions provided the following conditions are satisfied.

\medskip

\noindent
(i) The propagator powers and kinematic parameters obey
\begin{equation}
0 < \sum_i \nu_i - \frac{L D}{2} \le 1,
\end{equation}
where $L$ is the loop order, $D$ the spacetime dimension, and $\nu_i$ the propagator exponents.

\medskip

\noindent
(ii) The second Symanzik polynomial can be written in the form
\begin{equation}
F = A\, x + B,
\end{equation}
with $A \ge 0$ and $B \ge 0$.

\medskip

\noindent
Under these assumptions, the integral defines a Stieltjes function of the variable $x$ when all the other variables are held fixed in the Euclidean region. For planar Feynman integrals, this condition is satisfied quite generally: the variable $x$ may be taken to be any kinematic invariant of the form
\begin{equation}
x = -(p_i + p_{i+1} + \cdots + p_{j-1})^2 \quad \text{or} \quad x = m_i^2,
\end{equation}
with all other variables held fixed in the Euclidean region.

\medskip

\noindent
For non-planar integrals, there is strong evidence from explicit examples that the above condition on the Symanzik polynomial continues to hold, although a general proof remains an open problem.
\end{theorem}
\end{tcolorbox}

As we review in section~\ref{sec:numboot} these positivity properties can be used to numerically constrain Feynman integrals. Remarkably these properties allow us to constrain Feynman integrals not just in the Euclidean region but also in Lorentzian kinematics.

\subsection{Relation between dispersion relations and positivity}
In certain cases the existence of a dispersion relation and unitarity can imply positivity properties. In these cases we can imagine that the positivity properties are a non-trivial consequence of unitarity, causality/analyticity of the S-matrix.

Consider a function $A(s)$ that satisfies an \textit{unsubtracted} dispersion relation
\begin{equation}
    A(s) = \int_{4m^2}^{\infty} \, ds' \,\,\frac{\text{Disc } A(s')}{s' - s}
\end{equation}
By performing a variable transformation to the Euclidean region $x = -s$, and using the Schwinger trick 
\begin{equation}\label{eq:schwingertrick}
\frac{1}{x+s'}= \int_{0}^{\infty} dt\, e^{-t(x+s')}\quad
\end{equation}
\noindent we can rewrite the amplitude as a Laplace transform:
\begin{equation}
    A(x) = \int_{0}^{\infty} dt \, e^{-tx} \underbrace{\int_{4m^2}^{\infty} ds' \, e^{-ts'} \text{Disc } A(s')}_{\mu(t)}
\end{equation}
This representation allows us to classify the amplitude based on the nature of its spectral density:
\begin{itemize}
    \item \textbf{Stieltjes Property:} If $\text{Disc } A(s') \geq 0$, then $A(x)$ is a Stieltjes function. This is a \textbf{stronger} condition that can typically follow from Unitarity and the Optical Theorem.
    \item \textbf{Complete Monotonicity (CM):} If $\mu(t) \geq 0$, then $A(x)$ is a CM function for $x > 0$. 
\end{itemize}
\subsubsection{Spectral Representation of Vacuum Polarization}

The vacuum polarization function $\Pi(q^2)$ can be described using the \emph{K\"all\'en-Lehmann spectral representation}, which relates the full propagator to a spectral density $\rho(s)$. For gauge-invariant observables, the spectral density $\rho(s)$ is guaranteed to be positive-definite for all $s \ge s_{thr}$, i.e., above the threshold for particle production. This positivity is a direct consequence of the unitarity of the theory. To ensure the photon remains massless ($\Pi(0)=0$) and to render the integral ultraviolet finite, one usually employs a \textbf{once-subtracted dispersion relation}:
\begin{equation}
    \Pi(q^2) = q^2 \int_{s_{thr}}^{\infty} \frac{\rho(s)}{s(s - q^2)} ds
\end{equation}
By changing of variables $u = 1/s$ and $Q^2 = -q^2 > 0$, we get 
\begin{equation}
    \Pi(Q^2) = -Q^2 \int_{0}^{1/s_{thr}} \frac{\rho(1/u)}{1 + u\,  Q^2} du
\end{equation} 

\begin{tcolorbox}[colback=blue!5!white,colframe=blue!60!black,title=Stieltjes property of the vaccum polarization]
If $\Pi(Q^2)$ is the vaccum polarization function then $\frac{- \Pi(Q^2)}{Q^2}$ is a Stieltjes function.
\end{tcolorbox}

\begin{figure}[H]
    \centering
    \includegraphics[width=0.5\linewidth]{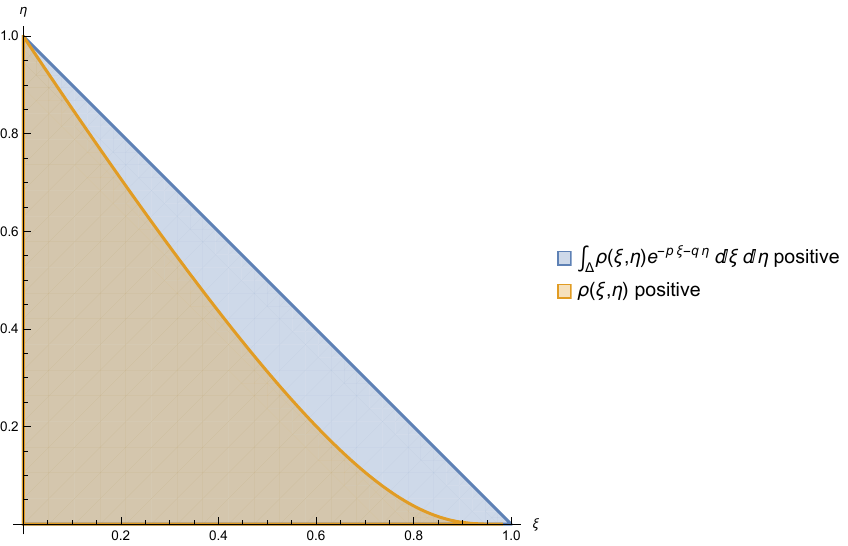}
    \caption{Regions inside the integration domain $\Delta$ where eq.(113),(114) hold fot $L=2$.}
    \label{fig:Coulombtriang}
\end{figure}
\subsection{Coulomb branch amplitudes in $\mathcal{N}=4$ SYM}
The Coulomb branch amplitudes \cite{Alday:2009zm}, which depend on the kinematic variables $u=4 m^2/(-s), v=4 m^2/(-t)$.

\begin{tcolorbox}[colback=blue!5!white,colframe=blue!60!black, title= Positivity properties of Coulomb branch amplitudes]
 $(-1)^L {\cal M}^{(L)}(u,v)$ is a CM function of $u,v$, for $u >0,v >0$ using the available results \cite{Caron-Huot:2014lda} at $L=1,2,3$.  
 
 For $L=1,\quad$  $-\mathcal{M}^{(1)}(u,v)$ is also a Stieltjes function. 
\end{tcolorbox}

The 4-point amplitudes on the Coulomb branch admit a remarkably simple Mandelstam representation \cite{Caron-Huot:2014lda}:
\begin{equation}
    M(u, v) = \int_{\Delta} d\xi d\eta \frac{\rho(\xi, \eta)}{(\xi + u)(\eta + v)}
\end{equation}
where $\Delta$ is the integration region $\Delta=\{\xi, \,\, \eta\, \,\ge 0,\, \, \xi + \eta \leq 1\}$. Just like in the previous section by using Schwinger's trick eq.~\eqref{eq:schwingertrick} the behavior of the double spectral function $\rho(\xi, \eta)$ determines the mathematical class to which of the amplitude:
\begin{itemize}
    \item \textbf{Stieltjes:} The amplitude is Stieltjes if \begin{equation}\label{eq:madelstramst}
       \rho(\xi, \eta) \geq 0. 
    \end{equation}
    \item  \textbf{Completely monotone:} The amplitude is CM for $u, v > 0$ if the following integral holds for all $p, q \geq 0$:
    \begin{equation}\label{eq:madelstramst1}
        \int_{\Delta} d\xi d\eta \, \rho(\xi, \eta) \, e^{-\xi p - \eta q} \geq 0
    \end{equation}

For $L=1$ eq.~\eqref{eq:madelstramst} holds but for $L=2$  only the weaker condition    eq.~\eqref{eq:madelstramst1} holds as shown in the figure~\ref{fig:Coulombtriang} 
\end{itemize}.


\begin{figure}[H]
    \centering
    \includegraphics[width=0.6\textwidth]{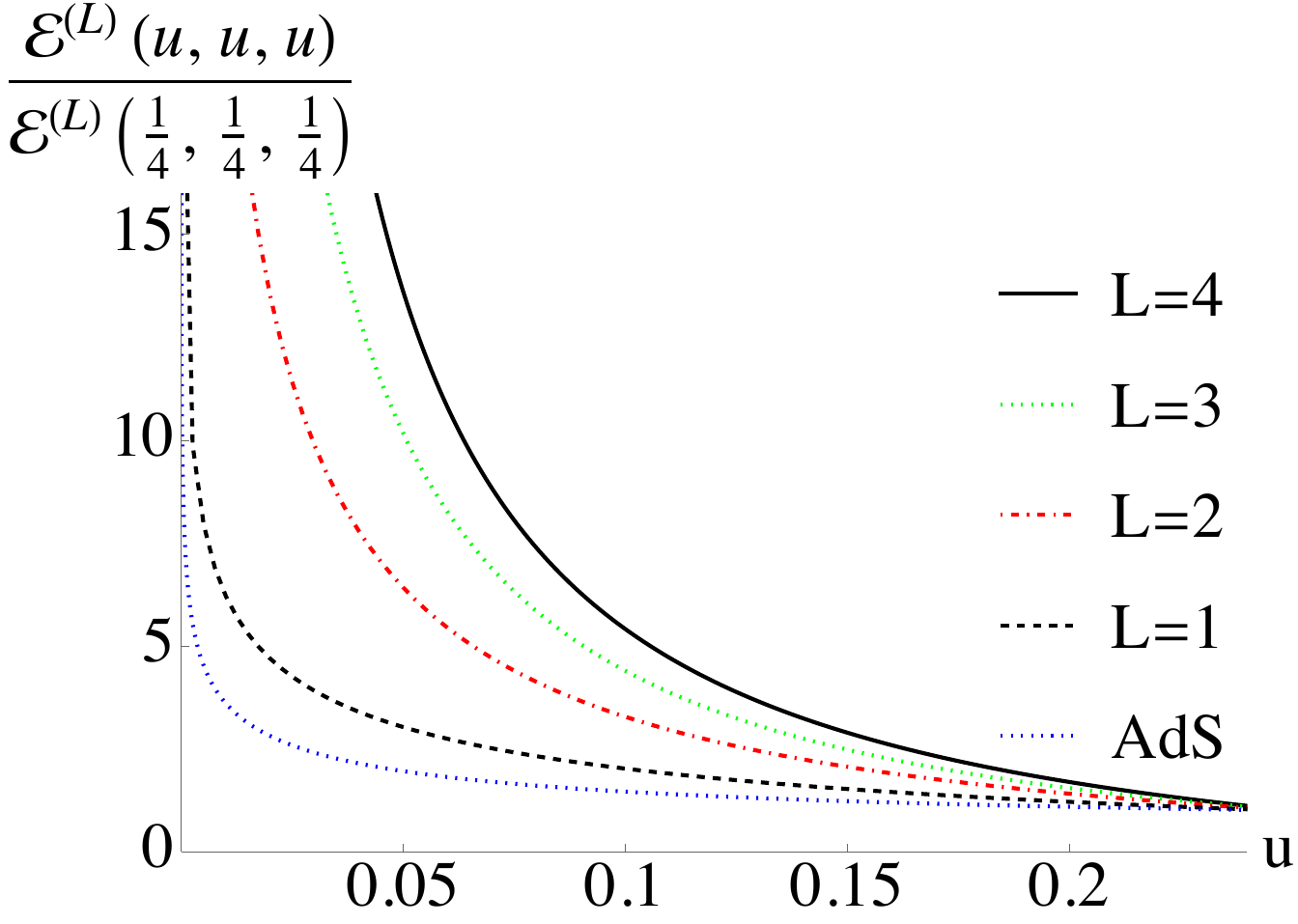}
    \caption{Behavior of ${\cal E}^{(L)}(u,u,u)$ in the region $0<u<1/4$. 
    From bottom to top: strong coupling and $L=1,2,3,4$ loop results. 
    The curves are positive, monotonically decreasing, and convex, as expected for completely monotone functions.}
    \label{fig:Efunction}
\end{figure}

\medskip 

In ref.~\cite{alday2025partonsstringsscatteringcoulomb}, the authors numerically bootstrap the four-point Coulomb branch amplitude in $\mathcal{N}=4$ SYM using S-matrix bootstrap techniques. The functional space is constrained by dual conformal invariance, while the numerical inputs are anchored by integrability-derived data and exact Regge trajectories across all coupling strengths. Remarkably, they find evidence that \textit{complete monotonicity} (CM) persists at finite coupling.

This result is particularly striking given that in the same ref.~\cite{alday2025partonsstringsscatteringcoulomb} it was demonstrated that the finite-coupling amplitude cannot satisfy a standard Mandelstam representation. Typically, CM is a consequence of the simple spectral properties associated with such representations, its persistence in their absence suggests that the positivity of the amplitude is governed by a deeper, perhaps more fundamental, geometric principle that transcends the traditional analytic structure of perturbative QFT. 

We will now discuss the connection to positive geometries which is indeed such a formulation of QFT. 

\medskip 

\subsection{Positive Geometries and dual volumes}\label{sec:posgeom}


The \emph{positive geometry program} associates certain physical theories with a \emph{positive geometry} and directly computes scattering amplitudes from its structure \cite{Arkani-Hamed:2013jha, Arkani-Hamed:2017tmz, Arkani-Hamed:2017mur}. Remarkably, in this framework, fundamental properties such as locality and unitarity emerge rather than being imposed. Tree-level amplitudes and loop-level integrands are encoded as \emph{differential forms} on the positive geometry.

\medskip
\noindent
\textbf{Examples:}
\begin{itemize}
    \item All loops, all multiplicities: planar $\mathcal{N}=4$ SYM, $\phi^3$ theory, ABJM theory.
    \item Up to one loop, all multiplicities: scalar theories with colour.
\end{itemize}

A $D$-dimensional positive geometry is a space carved out by a system of homogeneous polynomial equations and inequalities, with boundaries of all co-dimensions, and a unique associated meromorphic $D$-form, called the \emph{canonical form} \cite{Arkani-Hamed:2017tmz}. Every boundary component is itself a positive geometry of lower dimension, and taking the residue of the canonical form at a boundary gives the canonical form of that boundary.  

 At tree level, amplitudes can be interpreted as the \emph{volume of a dual polytope} \cite{Hodges:2009hk}. At loop level, this interpretation is conjectural: the \emph{dual amplituhedron} is not yet known \cite{Arkani-Hamed:2014dca}. While we do not yet know if the integrands do correspond to dual volumes they seem  to be positive inside the geometry. Integrating these forms produces scattering amplitudes, and while the positivity of the integrated results is non-trivial, empirical evidence supports it \cite{Dixon:2016apl}.

Following the framework of positive geometries (cf.\ ref.~\cite{Arkani-Hamed:2017tmz}), we can understand the origin of complete monotonicity as a consequence of the analytic structure of canonical forms.

\subsection*{Definitions and the Residue Property}
A \textit{positive geometry} is a pair $(X, X_{\ge 0})$ where $X$ is a complex projective variety and $X_{\ge 0}$ is a closed subset of the real part $X(\mathbb{R})$. It is equipped with a unique meromorphic $d$-form $\Omega(X, X_{\ge 0})$ called the \textit{canonical form}, uniquely determined by the property that it has simple poles only on the boundaries of $X_{\ge 0}$, and its residues are recursively defined:
\begin{equation}
    \text{Res}_{F} \Omega(X, X_{\ge 0}) = \Omega(F, F_{\ge 0})
\end{equation}
where $F$ is a boundary component (facet) of $X_{\ge 0}$. For a 0-dimensional point, $\Omega(\text{pt}) = \pm 1$.
\subsection*{Examples}
\begin{itemize}

\item[1.]{\bf The Interval:}
Consider the interval in $X=\mathbb{P}^{1}(\mathbb{R})$ with $X_{\ge 0} = \{ (1, x) \in \mathbb{P}^1(\mathbb{R}) \mid a \le x \le b \}$. 
The canonical form is the unique 1-form with simple poles at $a$ and $b$ with unit residues:
\begin{equation}
    \Omega([a, b]) = \left( \frac{1}{x-a} - \frac{1}{x-b} \right) dx = \frac{b-a}{(x-a)(x-b)} dx.
\end{equation}

\item[2.]{\bf The Triangle:}
Consider a triangle $T$ in $X=\mathbb{P}^2(\mathbb{R})$ with $X_{\ge 0} = \{ (1, x,y) \mid y \ge 0,\, 1-x-y\ge 0, \, 1+x-y\ge 0\}$. The canonical form takes the form:
\begin{equation}
    \Omega(T) = \frac{N \, dx \wedge dy}{y~(1-x-y)~(1+x-y)}
\end{equation}
Taking residues we see,
\begin{eqnarray}
  \text{Res}_{y=0,x=1} &=&\frac{N}{2}   \nonumber \\
  \text{Res}_{y=0,x=-1}&=&\frac{N}{2}\nonumber \\
  \text{Res}_{y=1,x=0} &=&\frac{-N}{2}
\end{eqnarray}
all of these have to be $\pm 1$ so we get $N=2$.
\item[3.]{\bf The Half-Pizza (Non-Polytopal Geometry)}:
Another example is \textit{half-pizza} geometry, which includes a non-linear boundary. It is defined by the region:
\begin{equation} \label{half-pizza}
    \mathcal{P} = \{ (1, x, y) \mid y+\tfrac{1}{2} \ge 0, \ 1-x^2-y^2 \ge 0 \}
\end{equation}
The canonical form for this region is:
\begin{equation}
    \Omega_{\mathcal{P}} = \frac{\sqrt{3} \, dx \wedge dy}{(y+\tfrac{1}{2}) ~(1-x^2-y^2)}
\end{equation}
\end{itemize}
\subsection*{Dual Laplace Representation and Complete Monotonicity}

\medskip 

 Consider the case of convex polytopes $A \in \mathcal{P}^m(\mathbb{R})$ in projective space, which in simple cases describe scattering amplitudes \cite{Hodges:2009hk}.
The canonical rational function  ${\underline \Omega(Y)}= \frac{\Omega(Y)}{\langle Y d^m Y\rangle}$ of such a polytope is the volume of the dual polytope 
$A^{*}_Y$. 
The latter is defined by the facet inequalities $W\cdot Y>0$, for $Y \in A$ and for any $W \in A^{*}_Y$. The canonical function admits the following Laplace representation \cite{Arkani-Hamed:2017tmz},
\begin{equation}
{\underline\Omega}(Y)=\frac{1}{m!} 
\int_{W \in A^{*}_Y}  e^{-W \cdot Y} d^{m+1}W
\,.
\end{equation}
This proves that ${\underline\Omega}(Y)$ is CM for any $Y \in A$ by Choquet's theorem eq.~\ref{eqn:choquet}, with measure $\mu=1$.
According to this theorem, the directional derivatives are to be taken along the extremal rays of the dual polytope $A^{*}_Y$.

\begin{tcolorbox}[colback=blue!5!white,colframe=blue!60!black, title= Canonical rational function of any convex polytope is completely monotone]
 The canonical rational function ${\underline\Omega}(Y)$ of any convex polytope $A \in \mathcal{P}^m(\mathbb{R})$  for any $Y\in A$ is completely monotone. 

\end{tcolorbox}

This fact suggests to us a close connection between the CM property and dual geometries.
This could help when looking for a dual geometry in cases beyond polytopes, such as the conjectured dual Amplituhedron \cite{Arkani-Hamed:2010wgm,Ferro:2015grk,Herrmann:2020qlt}.

\subsection{Complete monotonicity of six-particle amplitudes inside the Amplituhedron}
An important motivation for studying positivity properties of integrated observables comes from positive geometry. 
In planar $\mathcal{N}=4$ super Yang--Mills theory, scattering amplitudes are associated with geometric objects known as Amplituhedra. 
At the level of loop integrands, these objects lead to rational functions that are conjectured to admit a geometric (volume) interpretation and are manifestly positive inside the corresponding region~\cite{Hodges:2009hk,Arkani-Hamed:2010wgm,Arkani-Hamed:2013jha}.

A natural question is whether such positivity properties persist after integration over loop momenta. 
Evidence for this was found in~\cite{Arkani-Hamed:2014dca}, where it was observed that suitably defined finite parts of integrated amplitudes remain positive when evaluated on kinematic configurations inside the tree-level Amplituhedron.

A particularly well-studied example is the six-particle amplitude after infrared subtraction. 
Using the BDS-like normalization~\cite{Bern:2005iz,Drummond:2007au,Caron-Huot:2016owq}, one obtains a finite, dual-conformally invariant function ${\cal E}(u,v,w)$ depending on three cross ratios. 
The relevant kinematic domain is the tree-level MHV Amplituhedron region
\begin{align}
P_{\rm MHV}:\quad 
u>0,\quad v>0,\quad w>0,\quad u+v+w<1,\quad 
(u+v+w-1)^2<4uvw \, .
\end{align}

We expand ${\cal E}$ perturbatively in the coupling,
\begin{align}
{\cal E}(u,v,w) = \sum_{L\geq 1} g^{2L} {\cal E}^{(L)}(u,v,w)\,.
\end{align}

\medskip

\begin{tcolorbox}[colback=blue!5!white,colframe=blue!75!black,title={Complete monotonicity in the Amplituhedron}]
For planar $\mathcal{N}=4$ super Yang--Mills theory, consider the six-particle amplitude ${\cal E}(u,v,w)$ in the BDS-like normalization. 
Then, for kinematic variables $(u,v,w)$ inside the tree-level MHV Amplituhedron region $P_{\rm MHV}$, the perturbative coefficients satisfy
\begin{align}
(-1)^L {\cal E}^{(L)}(u,v,w) \ \text{is completely monotone in}\ (u,v,w)\,.
\end{align}
This property has been proven analytically at one and two loops, is supported by strong numerical evidence at three and four loops, and is consistent with the strong-coupling result.
\end{tcolorbox}

At low loop orders, this statement can be established rigorously. 
Complete monotonicity has been proven analytically for $L=1$ and $L=2$, using explicit representations of the amplitude in terms of polylogarithmic functions and Feynman integrals. 

At higher loop orders,
Strong numerical evidence supports the validity of complete monotonicity at $L=3$ and $L=4$ throughout the Amplituhedron region.

By restricting to the symmetric slice $u=v=w$, where the kinematic domain reduces to $0<u<1/4$. 
On this slice, the strong-coupling result derived from AdS/CFT can be shown to satisfy complete monotonicity as well. 
This provides a non-perturbative consistency check of the conjectured structure.

The qualitative behavior displayed in Fig.~\ref{fig:Efunction} is characteristic of completely monotone functions: 
the functions are positive, decreasing, and convex throughout the domain. 
Remarkably, this simple pattern emerges despite the highly non-trivial analytic structure of the underlying amplitudes, which involve iterated integrals of increasing transcendental weight. This example provides strong evidence that positivity properties inherited from the Amplituhedron extend beyond the level of integrands and persist, in a highly non-trivial way, after integration.
\subsection{Other examples of positivity}
We finally mention that in ref.~\cite{henn2025positivitypropertiesscatteringamplitudes} and subsequently evidence for complete monotonicity  has been found for several other objects including Wilson loops with a Lagrangian insertions~\cite{Abreu:2024yit, Chicherin:2024hes}, cosmological correlators~\cite{henn2019fourloopcuspanomalousdimension}, energy correlators~\cite{Jaarsma:2025tck} and string amplitudes~\cite{henn2025positivitypropertiesscatteringamplitudes}. 

While our entire review has been about studying positivity properties in the kinematics, a natural question to ask would be if such properties can also be seen in the coupling. This was investigated for several exact observables in 4 dimensional super Yang Mills theories in ref.~\cite{Haensch:2025positivity} and strong evidence and in some cases proofs were found that several of these observables satisfy both complete monotoncity and Stieltjes properties in the coupling.
\section{Applications of positivity properties}

Having reviewed both the mathematical framework for Stieltjes and completely monotone functions and various examples where such properties arise in physics. We now look at some applications of these ideas. We will primarily follow refs.~\cite{mazzucchelli2025canonicalformsdualvolumes, ditsch2026approximatingfeynmanintegralsusing}
\subsection{Analytic S-matrix}

\subsubsection{Martin positivity bounds}

We now discuss an application of positivity properties in constraining the S-matrix and deriving rigorous non-perturbative bounds. We follow the discussion of \cite{Martin:1970jsp}.

Consider the $2\to 2$ scattering amplitude $M(s,t,u)$ of identical massive scalar particles of mass $m$. Under general physical assumptions—unitarity, analyticity, crossing symmetry, and polynomial boundedness (e.g.\ the Froissart--Martin bound)—the amplitude satisfies a fixed-$t$ dispersion relation. 

In particular, for $-t_0 \le t \le 4m^2$, one can write a twice-subtracted dispersion relation
\begin{equation}\label{eq:dispersion2}
    M(s,t)=c(t)+ \frac{1}{\pi} \int_{4 m^2}^{\infty} ds'\,
    \left(\frac{s^2}{s'^{2}(s'-s)}+ \frac{u^2}{s'^{2}(s'-u)}\right) 
    A(s',t),
\end{equation}
where $A(s',t)= \mathrm{Im}\, M(s',t)$ and $u=4m^2-s-t$.

Unitarity implies positivity of the absorptive part,
\begin{equation}
    A(s',t) > 0 \,, \qquad s' \ge 4m^2, \quad 0 \le t \le 4m^2.
\end{equation}

\medskip

\begin{figure}[H]
    \centering
    \includegraphics[width=0.65\linewidth]{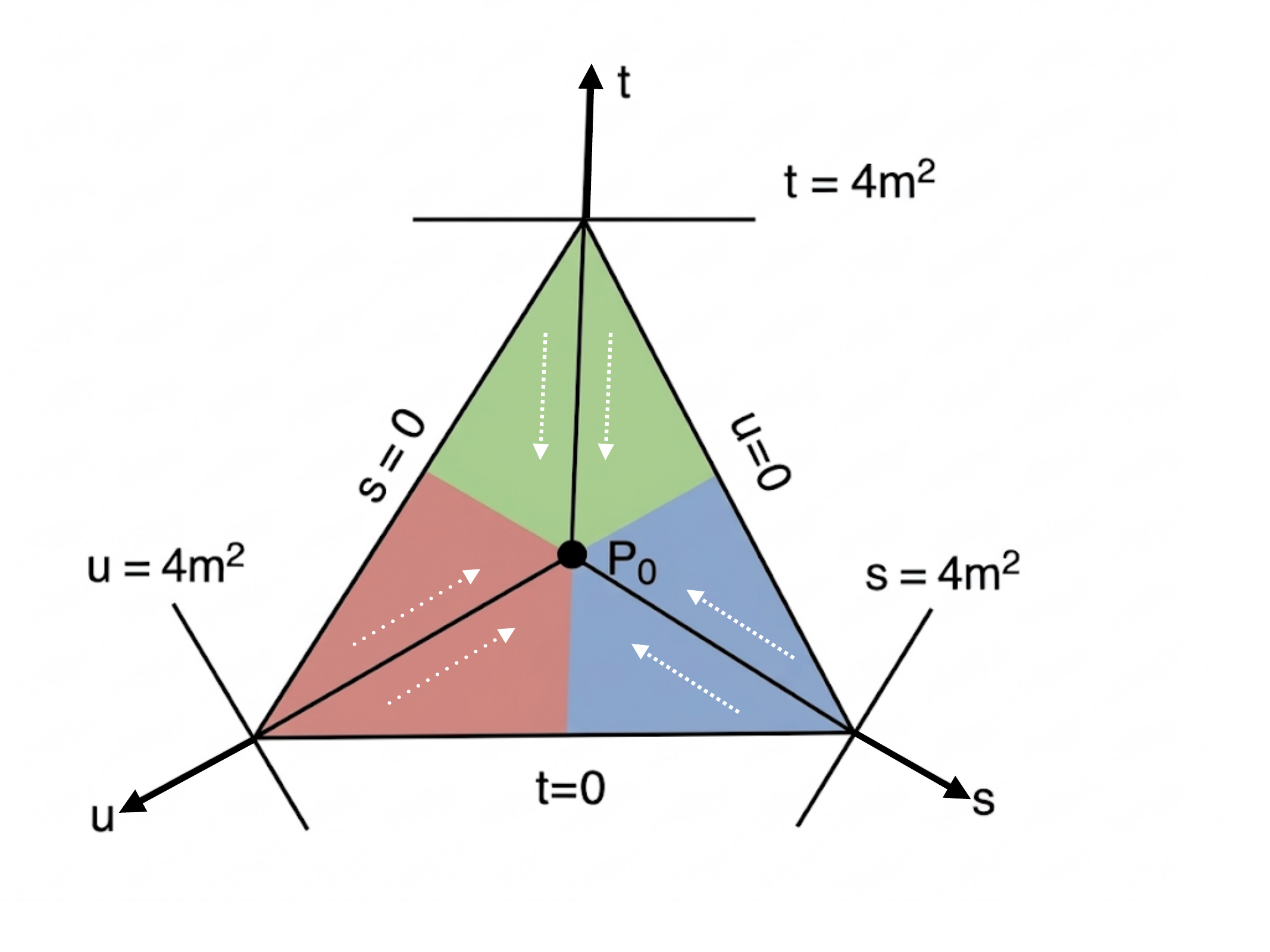}
    \caption{The Madelstram triangle for equal mass 2-2 sacttering. The coloured regions indicate where exactly one of the inequalities eq.~\eqref{eq:martin1}-\eqref{eq:martin2} are valid.}
    \label{fig:mantri}
\end{figure}

The Mandelstam triangle (the region $s,t,u < 4m^2$) corresponds to an unphysical region where the amplitude is real and constrained by crossing symmetry (see Figure~\ref{fig:mantri}). Its vertices lie at $s=4m^2$, $t=4m^2$, $u=4m^2$, while its edges correspond to $s=0$, $t=0$, $u=0$. The symmetric point 
\begin{equation}
P_0 = \tfrac{4m^2}{3}(1,1,1)
\end{equation}
is invariant under permutations of $(s,t,u)$.
\newpage
\vspace*{-20 pt}
\medskip
It can be shown \cite{Martin:1970jsp} that inside the Mandelstam triangle the amplitude satisfies the infinite set of positivity conditions, for all $n \ge 0$,
\begin{align}
\frac{d^n}{ds^n} M(s,t) &> 0 
&& \text{for fixed } t,\quad 2m^2 - \tfrac{t}{2} \le s \le 4 m^2, \label{eq:martin1} \\
\frac{d^n}{dt^n} M(s,t) &> 0 
&& \text{for fixed } s,\quad 2m^2 - \tfrac{s}{2} \le t \le 4 m^2, \label{eq:martin2} \\
\frac{d^n}{dt^n} M(4m^2 - u - t,\,t,\,u) &> 0 
&& \text{for fixed } u,\quad 2m^2 - \tfrac{u}{2} \le t \le 4 m^2. \label{eq:martin3}
\end{align}

\noindent
Up to a choice of conventions (in \cite{Martin:1970jsp} one uses $(s,t,u)\leftrightarrow (-s,-t,-u)$), these inequalities imply that the amplitude exhibits completely monotone behavior along each kinematic direction inside the Mandelstam triangle.

\medskip

These constraints lead to strong global properties of the amplitude. In particular, $M(s,t,u)$ attains its minimum at the crossing-symmetric point,
\begin{equation}
    M(s,t,u) \;\ge\; M\!\left(\tfrac{4m^2}{3},\,\tfrac{4m^2}{3},\,\tfrac{4m^2}{3}\right).
\end{equation}

We now give an intuitive argument for this fact. The Mandelstam triangle can be partitioned into three regions, in each of which one of the inequalities \eqref{eq:martin1}--\eqref{eq:martin3} controls the variation of the amplitude. Moving inward from any vertex toward the interior, the first-derivative positivity ($n=1$) implies that the amplitude decreases along these directions. The second-derivative condition ($n=2$) enforces convexity, and together with crossing symmetry ensures that the symmetric point $P_0$ is the unique global minimum.

\medskip

This provides a rigorous, non-perturbative lower bound for any scattering amplitude obeying \eqref{eq:dispersion2}. From the perspective of this work, this result can be understood as a direct consequence of the (directional) complete monotonicity of the amplitude inside the Mandelstam triangle. In fact, \cite{Martin:1970jsp} further uses these positivity properties to derive bounds on partial waves as well as absolute numerical bounds, for example
\begin{equation}
0 \le M\!\left(\tfrac{4}{3},\tfrac{4}{3},\tfrac{4}{3}\right) 
< M(2,2,0) < 3.6 \quad \text{for } m^2=1.
\end{equation}

\subsection{Positive geometries and dual volumes}

Let us now return to the question posed at the end of section~\ref{sec:posgeom}, namely whether complete monotonicity can be used to construct duals of non-polytopal positive geometries.

\medskip

\begin{figure}[H]
   \begin{subfigure}{.5\textwidth}
  \includegraphics[width=.8\linewidth]{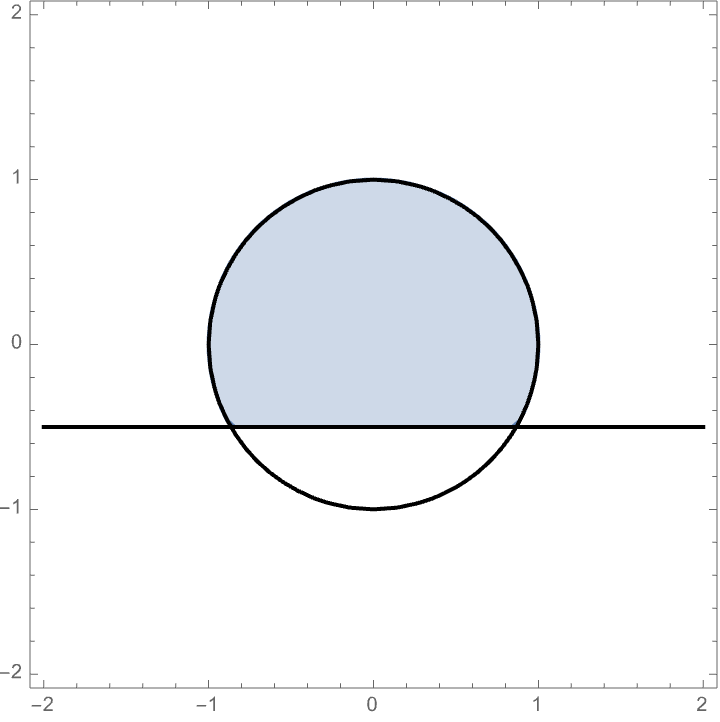}
  \caption{The half-pizza $\mathcal{P}$}
  \label{fig:sfig1}
\end{subfigure}%
\begin{subfigure}{.5\textwidth}
  \centering
  \includegraphics[width=.8\linewidth]{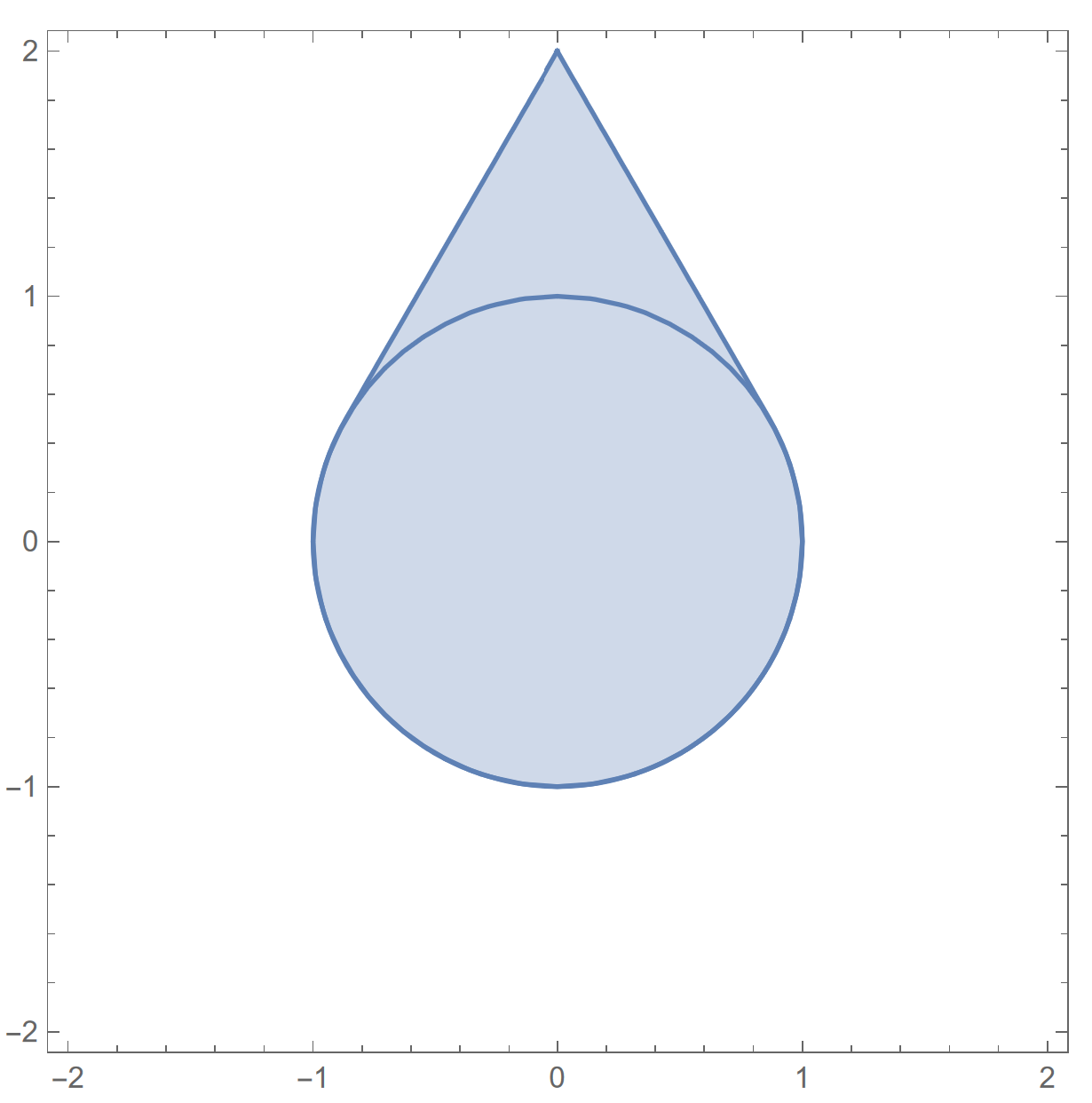}
  \caption{The dual half-pizza $\mathcal{P^*}$}
  \label{fig:sfig2}
\end{subfigure}
\label{fig:pizza}
\end{figure}

As a first example, consider the cone over the half-pizza geometry defined in eq.~\eqref{half-pizza}, given by the region (see Fig.~\ref{fig:sfig1})
\begin{equation}
    \mathcal{P} = \{ (z,x,y) \mid y+\tfrac{z}{2} \ge 0, \quad z^2 - x^2 - y^2 \ge 0 \}.
\end{equation}
Its canonical form is
\begin{equation}
    \Omega_{\mathcal{P}} = \frac{\sqrt{3}\, dx \wedge dy \wedge dz}{(y+\tfrac{z}{2})(z^2 - x^2 - y^2)}\,.
\end{equation}

\medskip

Using the construction of section~\eqref{sec:polardual}, the dual region is found to decompose as
\begin{equation}
    \mathcal{P}^* = \mathcal{P}_1 \cup \mathcal{P}_2 \,,
\end{equation}
where
\begin{align}
    \mathcal{P}_1 &= \{(w,u,v)\mid w \ge 0,\; w^2 - u^2 - v^2 \ge 0\}, \\
    \mathcal{P}_2 &= \{(w,u,v)\mid w \ge 0,\; |u|\le \tfrac{\sqrt{3}}{2} w,\; \sqrt{1-u^2} \le v \le 2 - \sqrt{3}|u| \}.
\end{align}

\medskip

We now attempt to reproduce the canonical function via a Laplace transform, as in the polytope case. Defining
\[
\underline{\Omega}_{\mathcal{P}} = \frac{\sqrt{3}}{(y+\tfrac{z}{2})(z^2 - x^2 - y^2)}\,,
\]
we compute
\begin{equation}
\int_{\mathcal{P}^*} du\, dv\, dw \; e^{-x u - y v - z w}
= \underline{\Omega}_{\mathcal{P}} + \underline{\Omega}_{\mathcal{T}}\,,
\end{equation}

where $\underline{\Omega}_{\mathcal{T}}$ is a transcendental contribution involving logarithms.

\medskip

Thus, unlike the polytope case, the naive dual volume does not reproduce the canonical function. If $\underline{\Omega}_{\mathcal{P}}$ admits an interpretation as a dual volume, this suggests two possibilities:
\begin{itemize}
    \item either $\mathcal{P}^*$ is not the correct notion of dual geometry, and the duality must be modified.
    \item or $\mathcal{P}^*$ is the correct dual, but the measure must be generalized to a non-trivial density $\mu(u,v,w)$.
\end{itemize}

\medskip

We can take inspiration from Bernstien--Hausdorff-Widder--Choquet theorem eq.~(\ref{eqn:choquet}) that  the latter is possible.  In ref.~\cite{mazzucchelli2025canonicalformsdualvolumes} we explored how to construct the non-trivial measures building on the works~\cite{Kozhasov:2019,Scott_2014}.

The question we just posed can be formalized with the notion of a Completely monotone positive geometry defined in~\cite{mazzucchelli2025canonicalformsdualvolumes}.

\begin{definition}[Completely monotone positive geometry]\label{def:CM_PG}
    We call a projective convex positive geometry $(\mathbb{P}^m,P)$ \textit{completely monotone} if the canonical function $\Omega_{\widehat{P}}$ is completely monotone on $\widehat{P}$, up to an overall choice of sign.
\end{definition}

It was already well known that rational functions which are completely monotone have a close relation with the notion of hyperbolic polynomials see refs.~\cite{Kozhasov:2019,Scott_2014}.
Let us now look at the definition and some examples of hyperbolic polynomials.
\begin{definition}[Hyperbolic polynomial]
A homogeneous polynomial $p(Y)$ on $\mathbb{R}^{m+1}$ is called \emph{hyperbolic} with respect to a vector $e \in \mathbb{R}^{m+1}$ if 
\[
p(e) > 0 \quad \text{and for every } Y \in \mathbb{R}^{m+1}, \text{ the univariate polynomial } t \mapsto p(Y - t\, e)
\]
has only real roots.
\end{definition}
Hyperbolicity means that any line in the direction $e$ intersects the hypersurface defined by $p(Y)=0$ in exactly $\deg(p)$ points (counted with multiplicities).
This property ensures that the \emph{hyperbolicity cone} 
\[
C_e = \{ Y \in \mathbb{R}^{m+1} : \text{all roots of } t \mapsto p(Y - t\, e) \text{ are non-negative} \}
\] 
is an open, convex cone. Moreover, if $p$ is hyperbolic with respect to $e$, it remains hyperbolic with respect to every vector in the connected component of $\mathbb{R}^{m+1} \setminus \{p=0\}$ that contains $e$.
\subsection*{Examples}

\begin{itemize}
    \item[\textbf{1.}] \textbf{Lorentz (light) cone:} The  polynomial $p(t,x,y) = t^2 - x^2 - y^2$ is hyperbolic with respect to $e=(1,0,0)$. Lines in the $t$-direction intersect the hypersurface $p=0$ in exactly 2 points. The hyperbolicity cone is the future light cone $t \ge \sqrt{x^2+y^2}$.

    \item[\textbf{2.}] \textbf{Smooth cubic curve:} The cubic polynomial
   $ p(x,y,z) = y^2 z - x^3 - 2 x z^2 - 3 z^3$ is hyperbolic with respect to $e=(0,1,1)$. Any line in the direction of $e$ intersects the cubic in exactly 3 points, and the hyperbolicity cone is the set of points from which moving along $-e$ meets only non-negative roots.

    \item[\textbf{3.}] \textbf{Positive semidefinite matrices:} 
    Let $X \in S^n$ be a symmetric $n \times n$ matrix. The determinant polynomial
    \[
    p(X) = \det(X)
    \]
    is hyperbolic with respect to the identity $e = I_n$. The hyperbolicity cone is the cone of positive definite matrices $S^n_+$.
\end{itemize}
It turns out a rational function can be completely monotone then its denominator has to be a hyperbolic polynomial see ref.~\cite{mazzucchelli2025canonicalformsdualvolumes}.
\begin{theorem}[Hyperbolicity from complete monotonicity]
Let $p, q \in \mathbb{R}[x_1, \dots, x_n]$ be homogeneous, coprime polynomials that are positive on an open convex cone $C \subset \mathbb{R}^n$. Suppose that, for some $\alpha > 0$, the function
\[ \label{Cmhyperboly}
f := \left(\frac{p}{q}\right)^{\alpha}
\]
is \emph{completely monotone} on $C$. Then $q$ is hyperbolic, and its hyperbolicity cone contains $C$.
\end{theorem}

\medskip

For us this means 
\begin{theorem}
If $(\mathbb{P}^m,P)$ is a completely monotone positive geometry, then $\partial_a P$ is cut out by a hyperbolic polynomial with hyperbolicity region equal to $P$.
\end{theorem}

So if the half-pizza $\mathcal{P}$ admits a dual volume representation via BHWC theorem then it has be a hyperbolic. Rather nicely this is indeed true and can be seen directly from figure~\ref{fig:sfig1} that if we choose any point inside $\mathcal{P}$ and draw a line passing through this point it intersects the curve at exactly 3 points. So this give us hope that a $\mu$ can exist. But how do we find it?

\medskip 

There are two approaches possible approaches.

\medskip
The first is related to the theory of hyperbolic PDE's with constant coefficients. The fundamental result is the following:

\begin{theorem}\label{thm:Riesz_p/q}
Let $p,q \in \mathbb{R}[x_1,\dots,x_n]$ be homogeneous polynomials, with $p$ hyperbolic and hyperbolicity cone $C$. Then
\begin{equation}
\frac{q(x)}{p(x)} = \int_{C^*} e^{-\langle x,y \rangle} \, \mu(y)\, dy \,, \quad \forall x \in C \,,
\end{equation}
where $\mu$ is a Schwartz distribution supported on $C^*$ given by
\begin{equation}\label{eq:fund_sol_int_repr_2}
\mu(y) = q(\partial)\,E(y) = (2\pi)^{-n} \int_{\mathbb{R}^n} e^{i\langle y,\xi\rangle}\, q(i\xi)\, p_{-}(\xi)^{-1}\, d\xi \,.
\end{equation}
\end{theorem}

\noindent
The expression $q(\partial)E(y)$ is understood in the distributional sense.

\medskip

The second approach is based on a geometric interpretation in terms of spectrahedral cones and their projections~\cite{Kozhasov:2019,Scott_2014}.

\begin{definition}[Spectrahedral shadow]
A \textit{spectrahedral shadow} is a linear projection of the cone of positive semidefinite matrices.
\end{definition}

\medskip

\noindent
\textbf{Idea.} Instead of working directly in $\mathbb{R}^n$, we embed our problem into the space of symmetric matrices, where positivity is easier to control, and then project back.

\medskip

Consider the polynomial
\begin{equation}\label{eq:spectr_pol}
p(x) = \det A(x) = \det(x_1 A_1 + \cdots + x_n A_n)\,.
\end{equation}

\medskip

The key fact is the following.

\begin{theorem}[Corollary 4.2 \cite{Scott_2014,Kozhasov:2019}]
Let $\alpha \in \{0,\tfrac{1}{2},1,\tfrac{3}{2},\dots,\tfrac{m-1}{2}\}$ or $\alpha > \tfrac{m-1}{2}$. Then $p^{-\alpha}$ is completely monotone on its spectrahedral cone $C$.
\end{theorem}

\medskip

\noindent
\textbf{Consequence.} For such $\alpha$, the function $p(x)^{-\alpha}$ admits a Laplace transform representation
\begin{equation}\label{eq:det_nu}
p(x)^{-\alpha} = \int_{S_m^*} e^{-{\rm tr}(A(x)B)} \, d\nu_\alpha(B)\,, 
\qquad x \in C\,,
\end{equation}
where $\nu_\alpha$ is a positive measure on the cone of positive semidefinite matrices $S_m^*$ (a version of the Wishart distribution).

\medskip

\noindent
\textbf{Geometric interpretation.} The dual variable $y$ does not correspond to a single matrix $B$, but to a whole family of matrices. This is encoded by a linear map
\[
L : S_m^* \to C^*\,.
\]
The associated Riesz measure is obtained by integrating over the corresponding fibers:
\begin{equation}\label{eq:Reisz_det}
\mu_\alpha(y) = \int_{L^{-1}(y)} d\nu_\alpha \,, 
\qquad y \in C^*\,.
\end{equation}

\medskip

\noindent
Thus, $\mu_\alpha(y)$ is obtained by integrating $\nu_\alpha$ over all matrices $B$ that project to $y$. Each fiber $L^{-1}(y)$ is itself a (possibly lower-dimensional) spectrahedron.

\noindent
\textbf{Summary.} The Laplace representation of $p(x)^{-\alpha}$ has a natural geometric meaning: the Riesz kernel $\mu_\alpha$ measures (possibly boundary) volumes of spectrahedral fibers in the dual cone.

\medskip

\noindent
\textbf{Useful integral representation.} For $\beta > \tfrac{m-1}{2}$ and any positive definite matrix $A$, one has
\begin{equation}\label{sh}
\frac{1}{(\det A)^{\beta}}=
\frac{1}{\pi^{m(m-1)/4} \prod_{j=0}^{m-1}\Gamma\left(\beta-\tfrac{j}{2} \right)}
\int_{B>0} e^{-{\rm tr}(A B)} (\det B)^{\beta-\tfrac{m+1}{2}} \, dB\,.
\end{equation}

\medskip

\noindent
To evaluate such integrals in practice, it is convenient to parametrize positive definite matrices as
\[
B = L L^T\,,
\]
where $L$ is lower triangular with positive diagonal entries. This reduces integration over matrices to integration over independent variables:
\begin{align} \label{lt}
\int_{B>0} f(B)\, dB 
= \int_{L^+} f(L L^T)\, J(L)\, dL\,,
\end{align}
where $J(L)$ is an explicit Jacobian and the variables of $L$ range over $\mathbb{R}$ (off-diagonal) and $(0,\infty)$ (diagonal).

\medskip

We now illustrate how this construction works in practice with a simple toy example.

\medskip

\paragraph{A warm-up example.}
Consider the function
\[
f(x,y,z)=\frac{1}{(z^2-x^2-y^2)^{\beta}}, \qquad \beta>\tfrac{1}{2}.
\]
Although this is not the canonical function of a positive geometry, it provides a useful example since the denominator
\[
p(z,x,y)=z^2-x^2-y^2
\]
is a hyperbolic polynomial.

Using the integral representation \eqref{sh} together with the parametrization \eqref{lt}, one obtains a Laplace transform representation of the form
\begin{equation}
\frac{1}{(z^2-x^2-y^2)^{\beta}}
= \int_{\bar{D}_2^{*}} e^{-x x'-y y'-z z'} \,\mu_{\beta}(x',y',z')\, dx' dy' dz',
\end{equation}
where $\bar{D}_2^{*}$ is the dual cone of the disk and
\[
\mu_{\beta}(x',y',z') = \big((z')^2-(x')^2-(y')^2\big)^{\beta-\tfrac{3}{2}}.
\]

\medskip

This example already illustrates an important point: the Laplace representation involves a \emph{non-trivial measure}. Only for the special value $\beta=\tfrac{3}{2}$ does the measure become constant, recovering the familiar polytope-like situation.

\medskip

\paragraph{The half-pizza geometry.}

We now return to the half-pizza geometry $\mathcal{P}$. In this case, the relevant matrix is
\[
A=\begin{pmatrix}
z-x & y & 0 \\
y & z+x & 0 \\
0 & 0 & y+\tfrac{z}{2}
\end{pmatrix}.
\]

Applying the same strategy as above leads to a Laplace transform representation over the dual region $\mathcal{P}^*$. After performing the change of variables and integrating out auxiliary parameters, one finds
\begin{equation}
\frac{1}{y(z^2-x^2-y^2)}
= \int_{\mathcal{P}^*} e^{-x x'-y y'-z z'} \,\mu(x',y',z')\, dx' dy' dz'.
\end{equation}

\medskip

The measure $\mu$ is given by
\begin{equation}
\mu(x',y',z')=
\begin{cases}
\displaystyle \frac{1}{2}+\frac{1}{\pi}\tan^{-1}\!\left(\frac{2y'-z'}{\sqrt{3}\,\sqrt{(z')^2-(x')^2-(y')^2}}\right),
& (x',y',z')\in \mathcal{P}_1, \\[1em]
1, & (x',y',z')\in \mathcal{P}_2.
\end{cases}
\end{equation}

\medskip

This measure is manifestly positive, since the arctangent takes values in $[-\pi,\pi]$.

\medskip

\paragraph{Key observation.}

This example highlights two important features:

\begin{itemize}
    \item The canonical function is recovered as a Laplace transform over the dual geometry $\mathcal{P}^*$, but only after introducing a \emph{non-trivial measure}.
    
    \item Unlike the polytope case, the measure is no longer algebraic: it involves \emph{transcendental functions} (in this case, an arctangent).
\end{itemize}

\medskip

This shows that even for relatively simple non-polytopal geometries, dual volume representations naturally require non-trivial and, in general, transcendental densities.

\medskip

Measures for several further examples, including geometries defined by intersections of lines, quadrics, and conics, were constructed in~\cite{mazzucchelli2025canonicalformsdualvolumes}.

\subsection{Numerical bootstrap applications}\label{sec:numboot}

A central theme of this review is that complete monotonicity (CM) and Stieltjes properties impose an infinite hierarchy of positivity constraints on a function and its derivatives. A natural question is whether these constraints can be turned into practical numerical tools for determining unknown functions.

The key observation is that, in many physical applications of interest, the functions under consideration are not arbitrary. In particular, Feynman integrals satisfy systems of differential equations, which drastically reduce the space of allowed functions. Concretely, if $f(x)$ satisfies a linear differential equation, then the set $\{f(x), f'(x), \ldots, f^{(k)}(x)\}$ spans a finite-dimensional vector space, and all higher derivatives can be expressed in terms of this basis. Equivalently, $f(x)$ may be part of a vector of master integrals
\begin{equation}
\vec{f}(x)=\{f_1(x)=f(x),f_2(x),\dots,f_n(x)\}\,, \qquad 
\partial_x \vec{f}(x) = A(x)\, \vec{f}(x)\,.
\end{equation}

This reduction from an infinite-dimensional function space to a finite-dimensional one is crucial. It allows us to combine differential equations with the infinitely many inequalities implied by CM and Stieltjes properties, thereby turning a qualitative structural property into a quantitative computational method.

\medskip

\noindent
\textbf{Convex optimization from positivity.} 
Once restricted to a finite-dimensional space, the CM and Stieltjes constraints can be implemented as convex optimization problems. The idea is to impose truncated versions of the positivity conditions and use them to bound the function and its derivatives at a given point.

\begin{itemize}
\item {\bf Linear programming:}

Imposing the truncated CM conditions
\begin{equation}
(-1)^n f^{(n)}(x)\geq 0\,, \qquad 0\le n\le n_{\max}\,,
\end{equation}
one can set up optimization problems such as
\begin{tcolorbox}
\begin{align}
{\rm Maximize / Minimize:} \quad 
\frac{(-1)^i f^{(i)}(x)}{f(x)}\bigg|_{x=x_0} \,,
\end{align}
subject to
\begin{align}
(-1)^n f^{(n)}(x)\geq 0\,, \quad 0\le n\le n_{\max}\,.
\end{align}
\end{tcolorbox}

This yields rigorous upper and lower bounds on the function and its derivatives. Increasing $n_{\max}$ systematically improves the bounds.

\item {\bf Semidefinite programming:}

A stronger implementation uses Hankel matrix positivity. Defining
\[
A_1(n)= \{ (-1)^{i+j} f^{(i+j)}(x) \}_{i,j=0}^n \,, \qquad 
A_2(n)= \{ (-1)^{i+j+1} f^{(i+j+1)}(x) \}_{i,j=0}^n \,,
\]
complete monotonicity implies that both matrices are positive semidefinite:
\begin{tcolorbox}
\begin{align}
{\rm Maximize / Minimize:} \quad 
\frac{(-1)^i f^{(i)}(x)}{f(x)}\bigg|_{x=x_0}\,,
\end{align}
subject to
\begin{align}
A_1(n)\succeq 0\,, \quad A_2(n)\succeq 0 \,, \qquad 0\le n\le n_{\max}\,.
\end{align}
\end{tcolorbox}

These semidefinite constraints encode an infinite set of nonlinear inequalities in a compact form and typically lead to stronger bounds than linear programming.
\end{itemize}

\medskip

\noindent
\textbf{Relation to other bootstrap methods.}
This strategy fits into a broader paradigm: combining recursion relations (or differential equations) with positivity constraints to bootstrap physical observables. This idea has been highly successful in a variety of contexts, including quantum mechanics, matrix models, and effective field theories.

\begin{center}
\begin{tabular}{ |c |c| c| }
\hline
 Example  & Recursion & Hankel Positivity \\ 
 \hline
&&\\
 QM systems & $\langle 0|[H,O]|0\rangle=0$ & Unitarity \\
 &&\\
 & $O=\sum c_i x^i$ & $\langle 0|OO^{\dagger}|0\rangle \succeq 0$ \\
 && \\
 \hline
 &&\\
 Matrix models & Loop equations & Unitarity     \\
 \hline
 &&\\ 
 EFTs & Crossing symmetry & EFT-hedron \\
 \hline
&&\\
 CM/Stieltjes functions & Differential eqs / moment recursion & $A_i(n)\succeq 0$ \\
 && \\
 \hline 
\end{tabular}
\end{center}

In the present context, this philosophy leads to a new approach to Feynman integrals, combining CM constraints with Pad\'e approximation techniques.

\medskip

\noindent
\textbf{Numerical evaluation of Feynman integrals.}
The practical implementation relies on two complementary ingredients:
\begin{itemize}
    \item Complete monotonicity, which provides strong constraints in the Euclidean region,
    \item The Stieltjes property, which enables efficient analytic continuation.
\end{itemize}

\subsubsection{CM bootstrap and the bubble example}

Combining complete monotonicity with differential equations allows one to
systematically generate linear inequalities of the form
\begin{equation}
Q_n(x)\cdot \vec{f}(x) \geq 0\,,
\end{equation}
where the vectors $Q_n(x)$ encode the derivatives of the basis functions in terms
of $\vec{f}(x)$ itself. 
The first two cases are given by
\begin{align}
    Q_0=\mathbf{1}\,, \quad Q_1(x)=-A \,, 
\end{align}
and higher derivative matrices are obtained recursively via
\begin{align} 
    Q_n(x)=- \partial_x Q_{n-1}(x)+ Q_{n-1}(x)Q_1(x) \,.
    \label{eq:recursiondericative}
\end{align}

As more constraints are included, this region shrinks, leading to increasingly precise bounds.

\paragraph{Example: massive bubble in $D=2$}

\begin{figure}[t]
    \centering
    \includegraphics[width=0.6\textwidth]{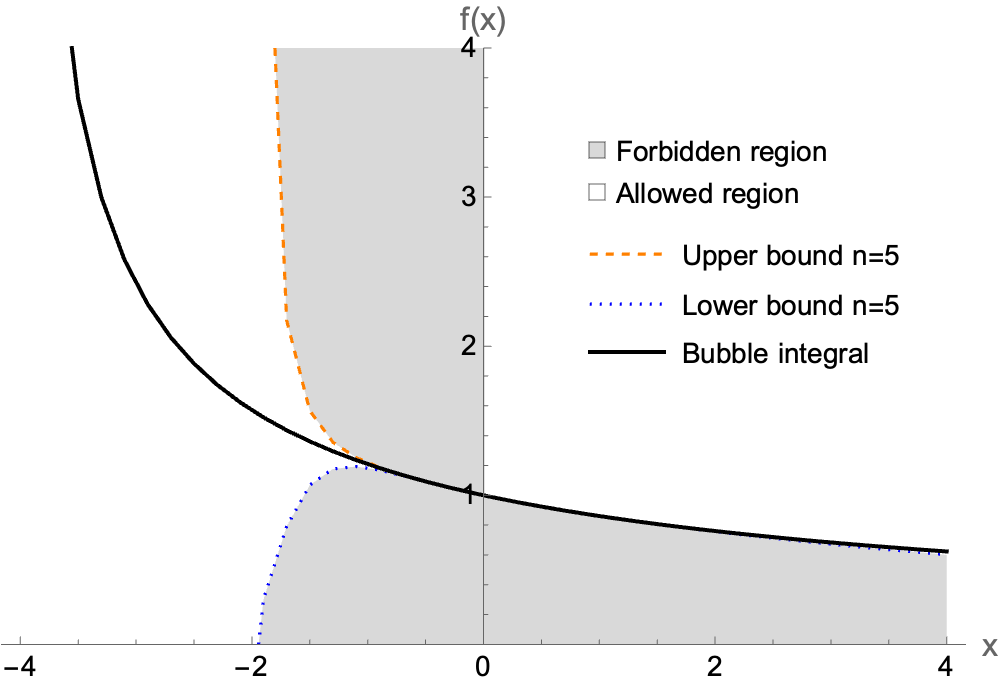}
    \caption{
    Bounds on the massive bubble integral obtained from truncated CM constraints taken from ref.~\cite{Ditsch:2025rdx}.
    }
 \label{fig:bubble_CM}
\end{figure}

As a simple example, consider the one-loop massive bubble integral
\begin{equation}
f(x) =
\int \frac{d^2k}{i\pi}
\frac{1}{(-k^2+m^2)\big(-(k+p)^2+m^2\big)} \,, 
\qquad x=-p^2 \,.
\end{equation}
The bubble family has two basis integrals, which we denote as 
\begin{align}
{\bf f} = \{  1, f\} \,.
\end{align}
The first integral is the tadpole integral, which depends on the mass $m$ only, and not on the parameter $x$. We therefore can fix its value to a constant. 
The differential equation for these basis integrals in $D=2-2~ \epsilon$ reads
\begin{align}\label{DEMatrices}
\partial_{x} {\bf f} = A_x \, {\bf f} \,, \quad  {\rm with} \quad
  A_{x} = -\begin{pmatrix}
0 & 0 \\
-\frac{2 }{(4 +x ) x} & \tfrac{2  +x+ ~\epsilon~ x}{(4 +x) x}  
\end{pmatrix}\,.
\end{align}
Applying the CM bootstrap yields upper and lower bounds on $f(x_0)$ which converge rapidly in regions where two-sided constraints exist. This is illustrated in figure.~\ref{fig:bubble_CM}.

This illustrates a key point: one can determine the value of a Feynman integral at a given Euclidean point without requiring its explicit analytic form.

\subsubsection{Stieltjes property and Pad\'e approximation}

While CM provides local control, it does not by itself allow one to efficiently move across kinematic regions. This is achieved by exploiting the Stieltjes property.

Expanding around a point $x_0$,
\begin{equation}
f(x) = \sum_{n=0}^\infty (-1)^n a_n (x-x_0)^n \,,
\end{equation}
one constructs Pad\'e approximants
\begin{equation}
P^N_M(x;x_0) = \frac{P_N(x)}{Q_M(x)} \,,
\end{equation}
which provide accurate approximations throughout the cut complex plane.

The resulting workflow is:
\begin{enumerate}
    \item Fix $f(x_0)$ using CM bootstrap,
    \item Compute derivatives via differential equations,
    \item Construct Pad\'e approximants,
    \item Evaluate them for general kinematics.
\end{enumerate}

\subsubsection{Example: 20-loop banana integral}

A particularly striking application is provided by equal-mass $L$-loop banana integrals in $D=2$, which admit the representation
\begin{equation}
I_L(x) = 2^L \int_0^\infty dt \, t\, J_0(t\sqrt{x})\, K_0(t)^{L+1} \,.
\end{equation}

Expanding around $x=0$ gives
\begin{equation}
f(x) = \sum_{n=0}^\infty (-1)^n a_n x^n \,,
\end{equation}
with coefficients given by moments of a positive measure.

Even at very high loop order (e.g.\ $L=20$), a relatively small number of moments suffices to construct highly accurate Pad\'e approximants, allowing efficient numerical evaluation without solving differential equations.

\begin{figure}[t]
    \centering
    \includegraphics[width=0.7\textwidth]{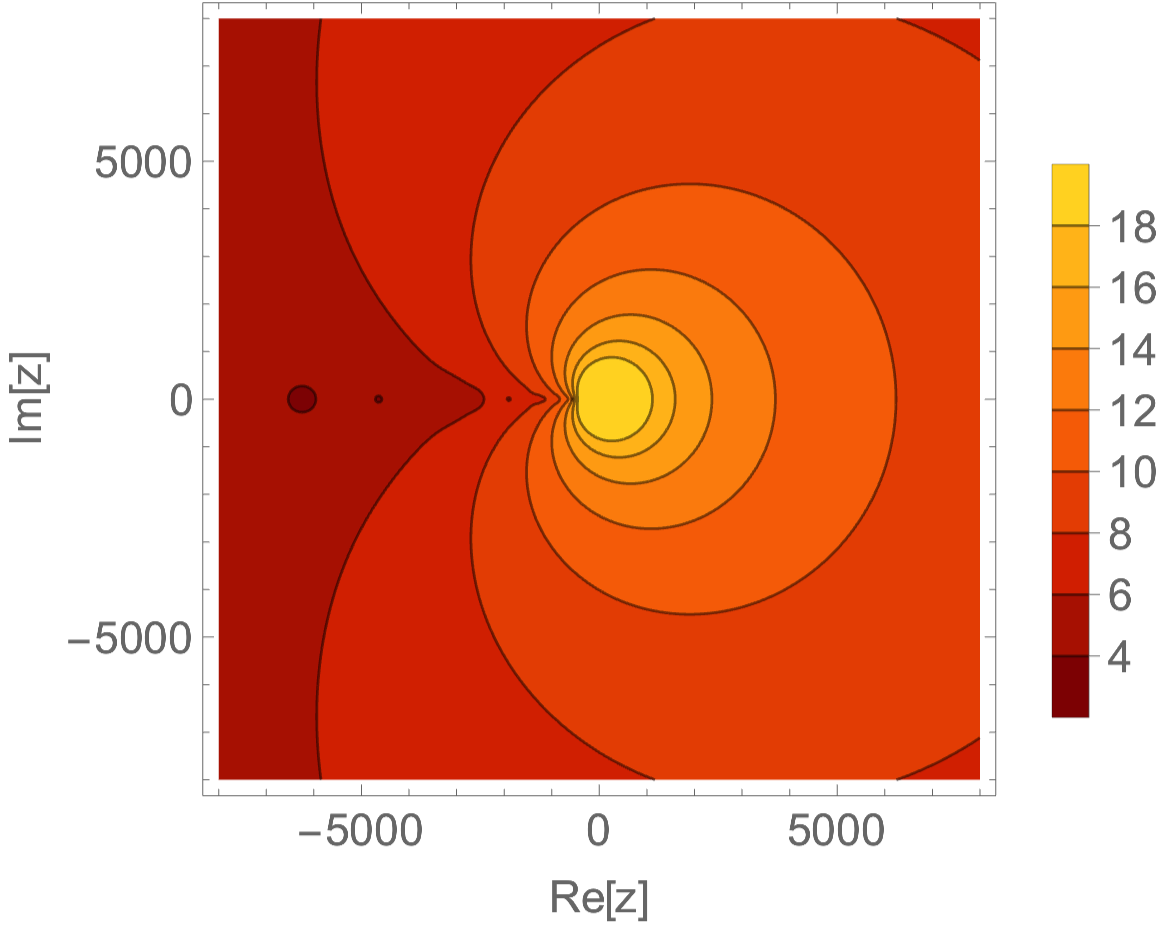}
    \caption{
    Numerical evaluation of the $20$-loop banana integral.
    Upper and lower Pad\'e approximants agree to high precision across a wide
    kinematic range, demonstrating excellent convergence.
    }
    \label{fig:banana_20loop}
\end{figure}

\medskip

\noindent
In summary, complete monotonicity provides rigorous local constraints in the Euclidean region, while the Stieltjes property enables controlled analytic continuation. Their combination leads to a powerful numerical bootstrap framework, turning general structural properties into efficient computational tools applicable from simple one-loop examples to very high-loop Feynman integrals.

\section{Summary and Outlook}

\subsection{Summary}

These notes have explored the interplay between complete monotonicity (CM), Stieltjes functions, and quantum field theory observables. A unifying theme is that many physically relevant quantities are governed by positivity and convexity, which in turn lead to integral representations with positive measures.

From a mathematical perspective, CM and Stieltjes functions provide a natural language for encoding these structures. Their defining properties imply strong constraints, such as infinite families of positivity conditions and analyticity in the complex plane. Conceptually, these properties reflect the fact that such functions can be built as superpositions of simple extremal building blocks.

On the physics side, we identified three main origins of these structures. First, parametric representations of Feynman integrals naturally give rise to CM and Stieltjes behavior in the Euclidean region. Second, unitarity and analyticity imply dispersion relations with positive spectral densities, leading directly to Stieltjes representations. Third, in theories with underlying positive geometry, canonical forms admit dual descriptions in which complete monotonicity follows from general convexity principles.

These insights have practical consequences. Positivity leads to rigorous bounds on observables, provides a foundation for numerical bootstrap methods, and ensures the stability and convergence of Pad\'e approximations for analytic continuation. More broadly, it suggests that a wide range of seemingly complicated functions in quantum field theory are controlled by a small set of structural principles.

\subsection{Outlook}

The results reviewed here raise a number of conceptual and technical questions.

A first set of problems concerns Feynman integrals. While positivity properties are well understood in many planar examples, their general validity—especially for non-planar graphs—remains unclear. Likewise, it would be desirable to identify intrinsic criteria that distinguish Stieltjes functions from more general completely monotone ones, and to clarify the correct multivariate generalization relevant for physical applications.

A second set of questions arises in the context of positive geometry. Although dual geometric interpretations are well understood in simple cases, their extension to more general and non-convex geometries, such as loop-level amplituhedra, is still largely open. In particular, the role and structure of the measures appearing in dual representations remain to be understood.

More broadly, there are intriguing indications that CM and Stieltjes properties persist beyond perturbation theory, for instance at finite coupling or in strongly coupled regimes. Explaining the origin of these structures—possibly in terms of deeper principles such as integrability or holography—would provide valuable insight into the nonperturbative organization of quantum field theory.

Finally, these ideas suggest a range of mathematical and phenomenological applications. On the mathematical side, connections to moment problems, approximation theory, and convex analysis deserve further exploration. On the physical side, it is natural to ask whether similar positivity structures appear in other settings, such as cosmological correlators, conformal field theory, or effective field theory.

\medskip

\noindent
Taken together, the recurring appearance of complete monotonicity and Stieltjes structure across geometry, analysis, and quantum field theory points to a common underlying principle: physically relevant observables are not arbitrary functions, but belong to highly constrained convex families. Understanding the origin and implications of this structure remains an important direction for future work.

\section*{Acknowledgements}

I am grateful to Sara Ditsch, Johannes Henn, Elia Mazzucchelli, and Maximilian Haensch for collaboration on the topics covered in these notes. I am particularly indebted to Johannes Henn for his constant encouragement and support.

It is a pleasure to thank Adolfo-Hilario Garcia, Leonardo de la Cruz, Sara Ditsch, Maximilian Haensch, Gregory Korchemsky, Jungwon Lim, Elia Mazzucchelli, Rainer Sinn, Bernd Sturmfels, Simon Telen, Jaroslav Trnka, Pierre Vanhove, Cristian Vergu, Qinglin Yang, Mao Zeng, Alexander Zhiboedov, and Simone Zoia for many helpful and stimulating discussions.

I also thank the participants of the \emph{Positive Geometry in Scattering Amplitudes and Cosmological Correlators} workshop (code: ICTS/PosG2025/02), held at the International Centre for Theoretical Sciences (ICTS), Bengaluru, in February 2025, where I had the opportunity to lecture on some of these topics. I am grateful as well to the Theory Group at CERN and to the Max Planck Institute for Mathematics in the Sciences for their hospitality during visits in which related material was presented.

I further thank the Erwin Schrödinger International Institute for Mathematics and Physics (ESI), University of Vienna, for its hospitality during the Thematic Programme \emph{Amplitudes and Algebraic Geometry} in 2026, where these notes were completed.

This work was funded by the European Union (ERC, UNIVERSE PLUS, 101118787). Views and opinions expressed are, however, those of the author(s) only and do not necessarily reflect those of the European Union or the European Research Council Executive Agency. Neither the European Union nor the granting authority can be held responsible for them.

\appendix
\section{Other positivity properties and integral representations}

A recurring theme in the study of positivity is that many physically relevant classes of functions form convex sets. A fundamental structural result explaining why such functions admit integral representations with positive kernels is \emph{Choquet's theorem}.

\medskip

\noindent
{\bf Theorem (Choquet).}
Let $K$ be a compact convex subset of a locally convex topological vector space. Then every point in $K$ can be represented as a barycenter (possibly continuous convex combination) of a probability measure supported on the extreme points of $K$.

\medskip

\noindent
In practical terms, this theorem states that any element of a convex set can be written as an integral over its extremal elements. When applied to spaces of functions obeying positivity constraints, this naturally leads to integral representations: functions are expressed as superpositions of extremal building blocks with respect to a positive measure.

\medskip

\noindent
This viewpoint underlies many classical representation theorems in analysis. In each case, positivity defines a convex cone of functions, and the corresponding extremal elements give rise to integral representations with non-negative measures.

\medskip

\noindent
Here we recall a few function classes that appear frequently in analysis and physics:
\begin{itemize}
    \item \textbf{Bernstein functions:} non-negative functions on $\mathbb{R}_{>0}$ whose derivative is completely monotone.
    \item \textbf{\(N\)-monotone functions:} functions on $(0,\infty)$ such that $(-1)^k f^{(k)}(x) \ge 0$ for $0 \le k \le N$.
    \item \textbf{Positive real functions:} analytic functions in a half-plane with non-negative real part.
\end{itemize}

\noindent
In all cases, the measure $\mu$ is non-negative, while the kernel reflects the analytic structure of the corresponding function class (Laplace, Stieltjes, or Cauchy-type kernels). The positivity of the measure encodes fundamental constraints, such as unitarity and causality in quantum field theory.

\medskip

\noindent
This general framework explains the ubiquity of integral representations with positive kernels: they arise naturally from convexity together with the characterization of functions in terms of their extreme points. 
\begin{center}
\renewcommand{\arraystretch}{1.5}
\setlength{\tabcolsep}{8pt}
\begin{tabular}{p{0.38\textwidth} p{0.58\textwidth}}
\toprule
\textbf{Function class} & \textbf{Integral representation (schematic)} \\
\midrule

Completely monotone on $\mathbb{R}_{>0}^n$ 
& $f(x)=\displaystyle \int_{\mathbb{R}_{\ge 0}^n} e^{-\langle x,t\rangle}\, d\mu(t)$ \\

Completely monotone on a cone $C$ 
& $f(x)=\displaystyle \int_{C^*} e^{-\langle x,t\rangle}\, d\mu(t)$ \\

Stieltjes functions 
& $f(x)=\displaystyle \int_{0}^{\infty} \frac{d\mu(t)}{1+t\,x}$ \\

Bernstein functions
& $f(x)=a+b\,x+\displaystyle \int_{0}^{\infty} (1-e^{-xt})\, d\mu(t)$ \\

Herglotz (Pick--Nevanlinna) functions 
& $f(z)=a+bz+\displaystyle \int_{\mathbb{R}} 
\left(\frac{1}{t-z}-\frac{t}{1+t^2}\right)\, d\mu(t)$ \\

$N$-monotone functions 
& $f(z)=\displaystyle \int_{0}^{1/z} (1-zt)^{N-1}\, dt$ \\

Positive real functions
& $f(z)=\displaystyle \int_{0}^{\infty} \frac{1+zt}{t-z}\, d\mu(t)$ \\

\bottomrule
\end{tabular}
\end{center}

\section{Moment problem}\label{app:momentproblem}
Let \( I \subset \mathbb{R} \) be an interval, and let \( \{s_n\}_{n=0}^{\infty} \) be a sequence of real numbers. The moment problem on \( I \) asks whether this sequence can be realized as the moments of a positive measure supported on \( I \), and if so, how uniquely this can be done.

\begin{enumerate}
    \item \textbf{Existence:} Can we find a positive Borel measure \( \mu \), supported on \( I \), such that
    \begin{equation}
    s_n = \int_I x^n \, d\mu(x), \quad \text{for all } n \ge 0?
    \end{equation}

    \item \textbf{Uniqueness:} If such a measure exists, is it uniquely determined by its moments?  
    If yes, the problem is \textit{determinate}; otherwise, it is \textit{indeterminate}, and one seeks to describe all measures with the same moments.
\end{enumerate}

If the moment problem is indeterminate, then it has infinitely many solutions. Indeed, if \( \mu_1 \neq \mu_2 \) are two measures with the same moments, then for any \( \lambda \in [0,1] \),
\begin{equation}
\mu = \lambda \mu_1 + (1-\lambda)\mu_2
\end{equation}
is also a solution. The nature of the problem depends strongly on the interval \( I \). For instance on infinite intervals, there may exist nonzero functions orthogonal to all polynomials, i.e.,
\begin{equation}
\int_I u^n f(u)\,du = 0 \quad \forall n \ge 0.
\end{equation}
For example, on \( (0,\infty) \):
\begin{equation}
f(u) = u^{-\log u} \, \sin(2\pi \log u),
\end{equation}
which illustrates why indeterminacy can occur.  For finite intervals, this cannot happen because polynomials are dense in the space of continuous functions on a closed interval, so a nonzero function cannot have all moments equal to zero.

The three classical cases correspond to when the interval is finite, semi-infinite, or the entire real line:

\begin{itemize}
    \item \textbf{Hausdorff moment problem} (\( I = (0,1) \)):  
    A sequence \( \{s_n\} \) is a moment sequence of a positive measure on \( (0,1) \) if and only if it is completely monotone:
    \begin{equation}
    (-1)^k \Delta^k s_n \ge 0 \quad \forall n,k \ge 0,
    \end{equation}
    where \( \Delta s_n = s_{n+1} - s_n \).  
    The problem is always determinate.

    \item \textbf{Stieltjes moment problem} (\( I = (0,\infty) \)):  
    Existence holds if the Hankel matrices
    \begin{equation}
    H_n = (s_{i+j})_{i,j=0}^n, \quad H_n^{(1)} = (s_{i+j+1})_{i,j=0}^n
    \end{equation}
    are positive semidefinite for all \( n \ge 0 \).  
    Determinacy is ensured by Carleman's condition:
    \begin{equation}
    \sum_{n=1}^{\infty} s_{2n}^{-1/(2n)} = \infty.
    \end{equation}

    \item \textbf{Hamburger moment problem} (\( I = (-\infty,\infty) \)):  
    Existence holds if the Hankel matrices
    \begin{equation}
    H_n = (s_{i+j})_{i,j=0}^n
    \end{equation}
    are positive semidefinite for all \( n \ge 0 \), and Carleman's condition again guarantees determinacy.
\end{itemize}

\subsection{Relation to completely monotone and Stieltjes functions}

These moments problems are intimately connected to completely monotone and Stieltjes functions. In particular, Completely monotone sequences are closely related to the Hausdorff moment problem ans Stieltjes functions are related to the Stieltjes moment problem.

\begin{definition}
A sequence $\{a_n\}_{n\ge 0}$ is called \emph{completely monotone} if
\begin{align}
(-1)^k \Delta^k a_n \ge 0, \quad \forall n\ge 0,\; k\ge 0,
\end{align}
where $\Delta$ is the forward difference operator, $\Delta a_n = a_{n+1} - a_n$.
\end{definition}

\begin{definition}
A sequence $\{a_n\}_{n\ge 0}$ is called \emph{minimal completely monotone} if it is completely monotone and ceases to be so when $a_0$ is replaced by any smaller value.
\end{definition}

Minimal CM sequences correspond to measures without atoms at $t=0$ and play a special role in interpolation problems, ensuring uniqueness of the interpolating CM function. 

A classical result of Hausdorff \cite{Hausdorff1921} states that any completely monotone sequence $\{a_n\}$ can be represented as moments of a positive measure $\mu$ supported on $[0,1]$:
\begin{align}
a_n = \int_0^1 t^n \, d\mu(t), \quad n\ge 0.
\end{align}

Through the change of variables $t = e^{-x}$, this is related to the Bernstein--Hausdorff--Widder (BHW) theorem (eq.~\eqref{eq:BHW}) for CM functions on $(0,\infty)$. In particular, if $f(x)$ is CM on $(0,\infty)$, then $\{f(n)\}_{n\ge0}$ is a CM sequence. The converse holds precisely for minimal sequences:

\begin{theorem}[Widder, \cite{Widder1931}]
There exists a completely monotone function $f(x)$ on $(0,\infty)$ such that
\begin{align}\label{eq:minimalCMcond}
a_n = f(n), \quad n\ge 0,
\end{align}
if and only if $\{a_n\}_{n\ge0}$ is a minimal completely monotone sequence.
\end{theorem}

Thus, the interpolation problem reduces to identifying when a sequence is minimally completely monotone. The following characterization provides a practical criterion:

\begin{theorem}[Theorem 7, \cite{WangIsmailBatirGuo2020}]
A sequence $\{a_n\}_{n\ge0}$ is minimal completely monotone if and only if $\{a_n\}_{n\ge1}$ is completely monotone, $a_0$ is finite, and
\begin{align}
a_0 = \sum_{i=0}^{\infty} (-1)^i \Delta^i a_i.
\end{align}
\end{theorem}
\section{Abelian and Tauberian Theorems for Laplace and Stieltjes Transforms} \label{app:tauberian}
Abelian and Tauberian theorems connect the asymptotics of a function with those of its integral transforms. 
Abelian theorems describe how the behavior of a function near zero or infinity determines the asymptotics of its transform, often giving full expansions, including logarithmic terms. 
Tauberian theorems, in contrast, allow one to infer the asymptotics of the original function or measure from the leading behavior of its transform, typically under positivity or monotonicity assumptions. 
Together, these results provide a powerful framework for analyzing Laplace and Stieltjes transforms. We collect the relevant theorems here from references~\cite{hardy1949divergent,bleistein1975asymptotic,bingham1989regular,wong2001asymptotic}
\medskip

\begin{definition}[Slowly varying function]
A measurable function \( L:(0,\infty)\to(0,\infty) \) is slowly varying at infinity if
\begin{equation}
\lim_{x\to\infty} \frac{L(\lambda x)}{L(x)} = 1 \quad \text{for all } \lambda>0.
\end{equation}
Typical examples include \( L(x)=\log x \) or \( L(x)=(\log x)^a \) with \( a\in\mathbb{R} \).
\end{definition}

\medskip

\begin{theorem}[Tauberian theorem for Laplace transforms]
Let \( f:[0,\infty)\to[0,\infty) \) be locally integrable and ultimately monotone, with Laplace transform
\begin{equation}
F(s) = \int_0^\infty e^{-st} f(t)\,dt.
\end{equation}
If, for some \( \alpha>-1 \) and slowly varying \( L \),
\begin{equation}
F(s) \sim \Gamma(\alpha+1)\, s^{-\alpha-1} L(1/s), \quad s\to 0^+,
\end{equation}
then
\begin{equation}
f(t) \sim t^\alpha L(t), \quad t\to\infty.
\end{equation}
\end{theorem}

\medskip

\begin{theorem}[Tauberian theorem for Stieltjes transforms]
Let \( \mu \) be a positive measure on \( [0,\infty) \) with Stieltjes transform
\begin{equation}
f(z) = \int_0^\infty \frac{d\mu(t)}{1+z\, t}.
\end{equation}
If, for some \( 0<\alpha<1 \) and slowly varying \( L \),
\begin{equation}
f(z) \sim C\, z^{-\alpha} L(1/z), \quad z\to 0^+,
\end{equation}
then
\begin{equation}
\mu([0,x]) \sim \frac{C}{\Gamma(\alpha+1)\Gamma(1-\alpha)}\, x^\alpha L(x), \quad x\to\infty.
\end{equation}
\end{theorem}

\medskip

\noindent
The roles of $s,t$ can be interchanged. These results show that under positivity and mild regularity, the small-\( s \) or small-\( z \) behavior of the transform determines the large-\( t \) or large-\( x \) asymptotics of the original function or measure and vice versa. The Tauberian theorems give only the leading asymptotic behavior, recovering a full expansion requires stronger assumptions.

\medskip
\noindent The corresponding Abelian theorems are as follows:
\begin{theorem}[Wong--Wyman expansion for Laplace transforms]
Let \( f \) be smooth near \( t=0 \) and admit a complete expansion
\begin{equation}
f(t) \sim \sum_{n=0}^{\infty}\sum_{k=0}^{K_n} a_{n,k}\, t^{\alpha+n} (\log t)^k, \quad t\to 0^+,
\end{equation}
with \( \alpha>-1 \). Then
\begin{equation}
F(s) = \int_0^\infty e^{-st} f(t)\, dt \sim \sum_{n=0}^{\infty} \sum_{k=0}^{K_n} a_{n,k}\, \frac{d^k}{d\alpha^k} \big[ \Gamma(\alpha+n+1) s^{-\alpha-n-1} \big], \quad s\to\infty.
\end{equation}
\end{theorem}

\medskip

\begin{theorem}[Abelian expansion for Stieltjes transforms]
Let \( \rho(t) \) be locally integrable with full expansion as \( t\to\infty \):
\begin{equation}
\rho(t) \sim \sum_{n=0}^{\infty}\sum_{k=0}^{K_n} c_{n,k}\, t^{\beta-n-1} (\log t)^k.
\end{equation}
Then
\begin{equation}
A(z) = \int_0^\infty \frac{\rho(t)}{t+z}\, dt \sim \sum_{n=0}^{\infty}\sum_{k=0}^{K_n} c_{n,k}\, \frac{d^k}{d\beta^k} \Big[\Gamma(\beta-n)\Gamma(1-\beta+n) z^{-\beta+n} \Big], \quad z\to 0^+.
\end{equation}
\end{theorem}
The roles of $s,t$ can be interchanged under mid regularity assumptions.

\bibliographystyle{utphys} 
\bibliography{ff1.bib} 
\end{document}